\documentclass[aps,pra,twocolumn,epsfigure,superscriptaddress,longbibliography,nofootinbib]{revtex4-1}
\pdfoutput=1
\usepackage{verbatim}
\usepackage{csquotes}
\usepackage{epsf,epsfig,amsmath,amssymb,dsfont}
\usepackage{amsthm,mathrsfs} 
\usepackage{graphicx} 
\usepackage{float}
\usepackage{slashed}
\usepackage{float}
\usepackage{hyperref} 
\usepackage[]{subfigure}
\usepackage{multirow}
\usepackage{mathtools}
\usepackage{color}
\usepackage[framemethod=TikZ]{mdframed}
\mdfdefinestyle{MyFrame}{%
    linecolor=blue,
    outerlinewidth=2pt,
    roundcorner=20pt,
    innertopmargin=\baselineskip,
    innerbottommargin=\baselineskip,
    innerrightmargin=20pt,
    innerleftmargin=20pt,
    backgroundcolor=gray!50!white}

\DeclareMathOperator{\sign}{sign}

\newcommand{\be}{\begin{equation}}
\newcommand{\ee}{\end{equation}}
\newcommand{\baln}{\begin{align}}
\newcommand{\ealn}{\end{align}}
\newcommand{\ben}{\begin{equation*}}
\newcommand{\een}{\end{equation*}}

\long\def\symbolfootnote[#1]#2{\begingroup%
\def\thefootnote{\fnsymbol{footnote}}\footnote[#1]{#2}\endgroup}

\newcommand{\ket}[1]{\left| {#1} \right\rangle}
\newcommand{\bra}[1]{\left\langle {#1}\right|}
\newcommand{\braket}[1]{\langle {#1} \rangle}

\usepackage{tikz}
\usetikzlibrary{calc}

\xdefinecolor{mycolor}{RGB}{62,96,111} 
\colorlet{bancolor}{mycolor}

\begin{document}
\title{Talbot--Lau effect beyond the point-particle approximation}

\author{Alessio Belenchia}\email{These authors contributed equally to this work}
\affiliation{Centre for Theoretical Atomic, Molecular, and Optical Physics, School of Mathematics and Physics, Queen's University, Belfast BT7 1NN, United Kingdom}
\author{Giulio Gasbarri}\email{These authors contributed equally to this work}
\affiliation{Department of Physics and Astronomy, University of Southampton, SO17 1BJ, UK}
\author{Rainer Kaltenbaek}
\affiliation{Faculty of Mathematics and Physics, University of Ljubljana,
Jadranska ulica 19, 1000 Ljubljana, Slovenia}
\affiliation{Institute for Quantum Optics and Quantum Information Vienna, Austrian
Academy of Sciences, Boltzmanngasse 3, 1090 Vienna, Austria}
\author{Hendrik Ulbricht}
\affiliation{Department of Physics and Astronomy, University of Southampton, SO17 1BJ, UK}
\author{Mauro Paternostro}
\affiliation{Centre for Theoretical Atomic, Molecular, and Optical Physics, School of Mathematics and Physics, Queen's University, Belfast BT7 1NN, United Kingdom}

\begin{abstract}
Recent progress in matter-wave interferometry aims to directly probe the quantum properties of matter on ever increasing scales. However, in order to perform interferometric experiments with massive mesoscopic objects, taking into account the constraints on the experimental set-ups, the point-like particle approximation needs to be cast aside. In this work, we consider near-field interferometry based on the Talbot effects with a single optical grating for large spherical particles beyond the point-particle approximation. We account for the suppression of the coherent grating effect and, at the same time, the enhancement of the decoherence effects due to scattering and absorption of grating photons. 
\end{abstract}
\maketitle


\section{Introduction}\label{I}
The experimental observation of quantum superpositions at the macroscopic level has proven a tall order due mainly to quantum decoherence effects. In this context, matter-wave interferometry, which directly probes the superposition principle of quantum mechanics, offers the possibility of the testing of quantum mechanics and modification thereof with increasingly larger objects~\cite{hornberger2012colloquium}. This paves the way to the characterization of the quantum-classical transition and potentially the investigation of possible modifications of quantum mechanics~\cite{torovs2018bounds,torovs2017colored,arndt2014testing,bateman2014near,kaltenbaek2016macroscopic} and the assessment of quantum spacetime effects~\cite{bassi2017gravitational,kaltenbaek2016macroscopic}.

Near-field interferometry with optical gratings~\cite{PhysRevLett.68.472,PhysRevA.56.R4365,arndt2014matter}, instead of material ones, is of particular interest for exploring the limits of quantum mechanics. In combination with current levitation and cooling techniques, it is the core of recent proposals for observing quantum superposition of increasingly large systems, most prominently macro-molecules~\cite{brezger2003concepts,brezger2002matter,nairz2003quantum,gerlich2011quantum} and nano-spheres~\cite{bateman2014near,kaltenbaek2016macroscopic}. However, all the current proposals employing near-field interferometry with optical gratings work in a regime where the system of interest has a linear dimension much smaller than the grating laser's wavelength, i.e., when the Rayleigh approximation holds true. Thus, in view of applying this technique to larger and larger objects it is crucial to sidestep the point-like approximation and account for the reduced coherent effect of the grating in combination with its enhanced decoherence effects.

In this work we take a step in this direction by applying the formalism developed in Ref.~\cite{pflanzer2012master} to account for the decoherence effect due to scattered grating photons on spherical particles. The reduced coherent effect of the grating is also considered, and the effect of absorbed photons is touched upon. We give some examples of interference patterns and quantum visibilities when realistic experimental parameters are considered.  

The remainder of the paper is organized as follow. In Sec.~\ref{II} we briefly review the Talbot--Lau effect for optical grating and give the expressions for the interference pattern and Talbot coefficients which determine it. In Sec.~\ref{III}, we discuss the reduced coherent effect of the grating when the Rayleigh approximation is not valid. In Sec.~\ref{IV}, we introduce decoherence effects due to scattering of grating photons off spherical particles. Furthermore, we consider how to include the effect of absorption in the picture, whilst the fact that a fully fledged quantum formalism for this is currently not available. Finally, we discuss how to obtain the classical limit and its relevance to the problem at hand. In Sec.~\ref{V}, some examples of interference patterns are shown which highlight the effects of misusing the Rayleigh approximation. We conclude this work in Sec.~\ref{VI} with a discussion of the results and future perspectives. 

\section{Talbot--Lau effect for optical gratings}\label{II}

Here, we  provide a concise review of the Talbot effect for matter-wave interferometry in the eikonal approximation. A more in-depth analysis can be found in Refs.~\cite{hornberger2009theory,PhysRevA.70.053608,nimmrichter2014macroscopic,PhysRevA.78.023612}.

The dynamics of a polarizable quantum particle interacting with an electromagnetic standing wave in the interaction picture is described by the master equation
\begin{align}\label{eq:me}
\partial_{t}\rho_{t} = -\frac{i}{\hbar}[V(\hat{r}_{t},t),\rho_{t}] +\mathcal{L}_{t}(\rho_{t}),
\end{align}
where 
$V(\hat{r}_{t},t)$ is the interaction potential and $\mathcal{L}_{t}=\mathcal{L}_{\rm{sca}}+\mathcal{L}_{\rm{abs}}$ is the dissipative term taking into account the effects due to scattering $(\mathcal{L}_{\rm{sca}}$) and absorption ($\mathcal{L}_{\rm{abs}}$) of grating photons.

As a full quantum description of matter-light interaction encompassing both scattering and absorption mechanisms is currently lacking, we make use of the results in Ref.~\cite{pflanzer2012master} to describe both the coherent effects of the grating and the decoherence due to photon scattering, while making use of semiclassical arguments~\cite{nimmrichter2014macroscopic,PhysRevA.70.053608}, to include decoherence effects due to absorption. This results in the interaction potential 
\begin{align}
V(\hat{r}_{t},t) = -\frac{\epsilon_{0} \epsilon_{c}^{{\scriptscriptstyle R}}}{4}\int_{V_{n}(\hat{r}_{t})}d{\bf r}|E_{\rm{sw}}({\bf r})|^2,
\end{align}
where $\epsilon_0$ is the vacuum permittivity, $\epsilon_c^{{\scriptscriptstyle R}}$ is the real part of $\epsilon_{c}=3(\epsilon-1)/(\epsilon+2)$ with $\epsilon$ the relative permittivity, and $E_{\rm{sw}}({\bf r})$ is the electric field of the standing wave. The integral is extended over the volume $V_{n}(\hat{r})$ of the dielectric particle. We  give explicit forms for the scattering and absorption decoherence terms described by $\mathcal{L}_{\rm{sca}}(\rho)$ and $\mathcal{L}_{\rm{abs}}(\rho)$ in Sec.~\ref{III} [cf. Eqs.~\eqref{eq:l_scatt} and~\eqref{abs}, respectively].

The net effect of the optical grating on the matter-wave density matrix is then given by 
\begin{align}\label{eq:scattering}
\rho \to  \mathcal{T}e^{\int_{0}^{\tau_{int}} d\tau \mathfrak{L}_{\tau}}\rho,
\end{align}
where the super-operator $\mathfrak{L}_{t}$ is defined through $\mathfrak{L}_{t}(\rho) =-\frac{i}{\hbar}[V(\hat{x},t),\rho_{t}] +\mathcal{L}_{t}(\rho_{t})$, $\tau_{int}$ is the interaction time, and $\mathcal{T}$ denotes the time ordering operator. Note that, in the case of pulsed-light grating, the interaction time is determined by the duration of the pulse.

\begin{figure}[t!]
\centering
\includegraphics[width=0.8\columnwidth]{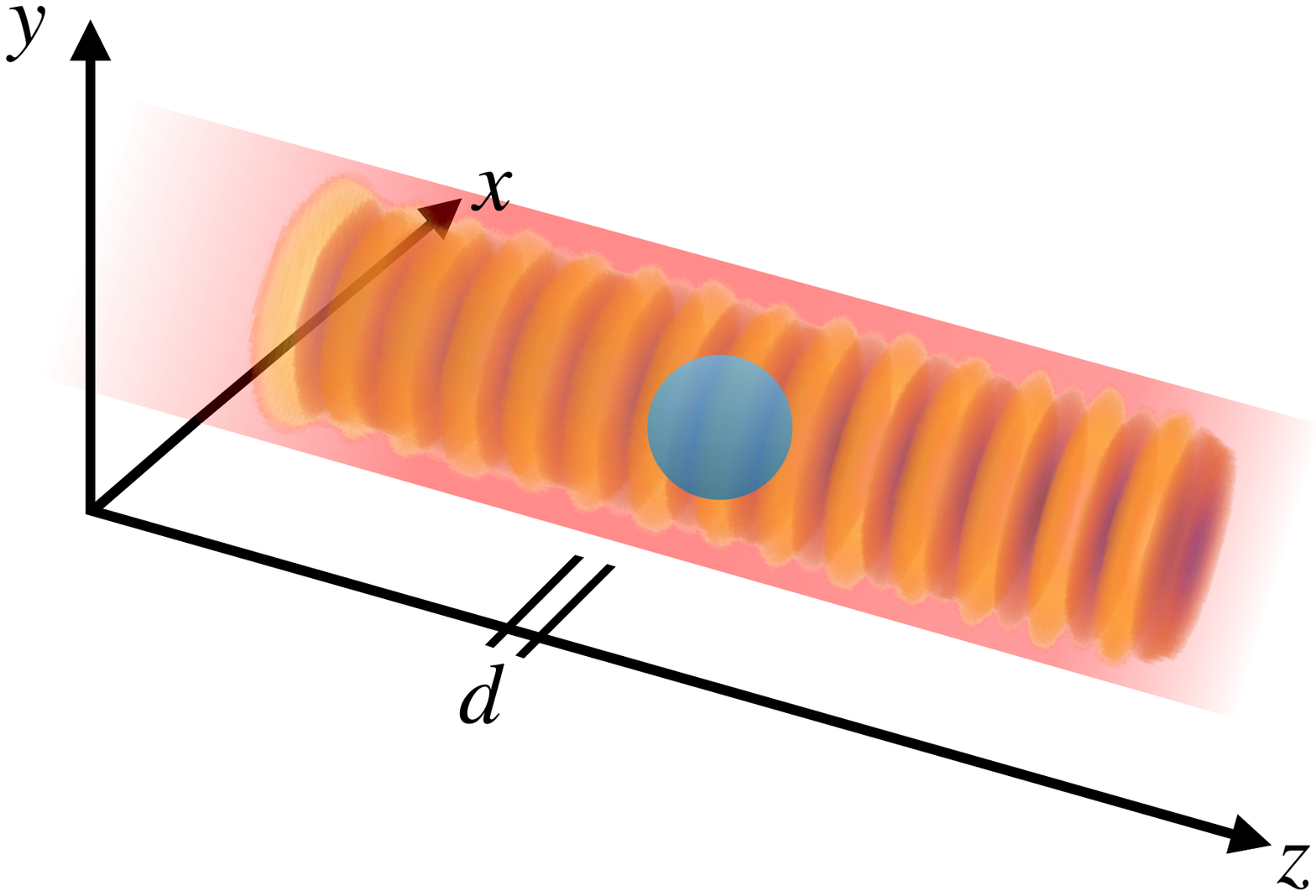}
  \caption{Optical standing-wave grating. The figure shows the coordinate convention that we use, in which $z$ is the relevant direction in the longitudinal eikonal approximation. We consider a standing-wave linearly polarized in the $x$ direction which gives rise to a grating of period $d=\lambda/2$.}
   \label{Fig:set-up}
\end{figure}

By assuming a waist of the standing wave that is much larger than the matter-wave profile, and an interaction time ($\tau_{int}$) that is  negligible compared to the characteristic time of spreading of the matter waves, we can rely on the longitudinal eikonal approximation~\cite{PhysRevA.78.023612} to reduce the problem to an effective one-dimensional dynamic along the direction of propagation of the standing-wave (cf. Fig.~\ref{Fig:set-up} for the coordinates setting).
This approximation allows us to rewrite Eq.~\eqref{eq:scattering} as
\begin{align}\label{eq:trgrat}
\rho(z,z')\rightarrow R(z,z')T(z,z')\rho(z,z'),
\end{align}
where $\rho(z,z') = \bra{z}\rho\ket{z'}$ is the matter-wave density matrix in the position representation, and we have introduced the phase-modification mask 
\begin{align}\label{Tco}
T(z,z')=t(z)t^{*}(z')=e^{-\frac{i}{\hbar}\int_{0}^{t}d\tau [V(z,\tau)-V(z',\tau)]}
\end{align}
with $V(z,t)$ the classical interaction potential, and a decoherence mask
\begin{align}\label{Rdeco}
R\left(z,z'\right)= e^{\int_{0}^{\tau_{int}} d\tau \mathcal{L}_{t}(z,z')}=R_{\rm{sca}}(z,z')R_{\rm{abs}}(z,z').
\end{align}



In order to make the description more transparent, it is convenient to rewrite Eq.~\eqref{eq:trgrat} in a phase-space picture.
We thus introduce the Wigner function associated with $\rho$ defined as
\begin{align}
w(z,p)=\frac{1}{2\pi\hbar}\int ds\, e^{\frac{i}{\hbar}ps}\bra{z-s/2}\rho\ket{z+s/2},	
\end{align}
with $z$ and $p$ the position and momentum coordinates in phase space. The effect of the grating in Eq.~\eqref{eq:trgrat} is then described by the action of a convolution kernel $\tilde{T}(z,p-q)$ on the Wigner function of the system~\cite{hornberger2009theory}, that is
\begin{align}
w'(z,p) = \int dq\, \tilde{T}(z,p-q)w(z,q).
\end{align}
The convolution kernel can be written explicitly as
\begin{align}\label{eq:conv}
\tilde{T}(z,p) = \int dq\, \mathcal{R}(z,p-q)\mathcal{T}_{\rm{coh}}(z, q),
\end{align}
where
\begin{align}\label{eq:conv_kern_tr}
\mathcal{T}_{\rm{coh}}(z,p) &= \frac{1}{2\pi \hbar}\int ds\,e^{\frac{ips}{\hbar}}t\left(z-\frac{s}{2}\right)t^{*}\left(z+\frac{s}{2}\right)
\end{align}
describes the coherent effect of the grating on the matter-wave, and
\begin{align}\label{eq:conv_kern_deco}
\mathcal{R}(z,p) &=\frac{1}{2\pi \hbar} \int ds\, e^{\frac{i p s}{\hbar}}R\left(z-\frac{s}{2},z+\frac{s}{2}\right)
\end{align}
accounts for the incoherent effects of the grating.
Exploiting the periodicity of the grating, the transmission function [Eq.~\eqref{Tco}] and  decoherence mask [Eq.~\eqref{Rdeco}] can be written in Fourier series as
\begin{align}\label{eq:fourier_co}
&t(z){=}\sum_{n=-\infty}^{\infty} b_{n}e^{\frac{2 \pi i n z}{d}},\nonumber\\
&R\left(z{-}\frac{s}{2},z{+}\frac{s}{2}\right) {=} \sum_{n=-\infty}^{\infty} R_{n}\left(\frac{s}{d}\right)e^{\frac{2 \pi i n z}{d}},
\end{align}
where $d$ is the grating period,$b_{n}=\frac{1}{d} \int_{-d/2}^{d/2} dz\, e^{\frac{2 \pi i n z}{d}} t(z)$, and 
\begin{align}
R_{n}\left(\frac{s}{d}\right)= \frac{1}{d} \int_{-d/2}^{d/2}\, dz e^{\frac{2 \pi i n z}{d}}R\left(z-\frac{s}{2},z+\frac{s}{2}\right).
\end{align}
Finally, using Eq.~\eqref{eq:fourier_co} the convolution kernel takes the form
\begin{align}\label{eq:conv_tb}
\tilde{T}(z,p) = \frac{1}{2\pi\hbar}\sum_{n} e^{\frac{2\pi i n z}{d}}\int \!ds\, e^{\frac{i}{\hbar}ps}  \tilde{B}_{n}\left(\frac{s}{d}\right)
\end{align}
where
\begin{align}\label{eq:TalbotCoeff}
\tilde{B}_{n}\left(\frac{s}{d}\right)=\sum_{j}B_{n-j}\left(\frac{s}{d}\right)R_{j}\left(\frac{s}{d}\right).
\end{align}
The $\tilde{B}_{n}$'s are known as Talbot coefficients and characterize the fringe pattern due to quantum matter-wave interference\footnote{In order to arrive at the full-fledged interference pattern, the evolution in phase space of the Wigner function from the source to the grating and from past the grating to the detection stage has to be obtained. Once the final Wigner function is given, the interference pattern can be straightforwardly derived noting that the position probability density function is just a marginal of the Wigner pseudo-probability. See e.g.~\cite{case2008wigner}}. These coefficients are conveniently expressed in Eq.~\eqref{eq:TalbotCoeff} in terms of the Fourier coefficients of the decoherence mask function $R_{n}$ and of the transmission function $b_{n}$. The latter are indeed contained in the Talbot coefficients $B_n$
\begin{align}
B_{n}\left(\frac{s}{d}\right)= \sum_{k}b_{k}b^{*}_{k-n}e^{\frac{i \pi (n-2k)s}{d}},
\end{align}
which describe only the coherent grating effect.

\section{Coherent grating for large spheres}\label{III}

The periodic modulation of the phase of the matter-wave quantum state of a polarizable particle operated by the optical grating is at the basis of the Talbot effect. 
We consider the case in which the grating is realized by retro-reflection of a laser pulse  off a mirror. This produces a linearly polarized standing wave field $\mathbf{E}(\mathbf{r})=E_0 \,\mathbf{\hat{e}}_x f(x,y)\cos(kz),$
where $f(x,y)$ is the transverse mode profile and $\mathbf{\hat{e}}_x$ is the polarization unit vector. In the following, we assume that the dimension of the particle is much smaller than the waist of the laser in the transverse directions. This allows us to neglect the transverse mode profile, and thus take $f(x,y)\simeq 1$.

The use of short laser pulses to generate the optical grating~\cite{nimmrichter2014macroscopic} justifies the eikonal approximation used to determine the coherent phase modulation and allows us to neglect any transverse force. Thus, we concentrate only on the reduced one-dimensional state of the matter wave along the standing-wave axis, i.e., we work in the longitudinal eikonal approximation. It should be noted that, set-ups with laser pulses have been used in~\cite{haslinger2013universal} and advocated for ground- and space-based experiments aiming to use massive objects~\cite{bateman2014near,kaltenbaek2016macroscopic}. On the contrary, the use of a continuous laser for the grating introduces limitations on the speed of the particles, which need to traverse the grating rapidly enough for the eikonal approximation to be valid. See~\cite{PhysRevA.78.023612} for how to go beyond the eikonal approximation.

\subsection{Polarizable point-like particles}
Let us start by reviewing well-known results on the effect of the optical grating for particles in the Rayleigh regime. Given $k=2\pi/\lambda$, the wave number of the standing-wave, and the radius $R$ of the particle, the condition for the particle to be in the Rayleigh regime reads $kR\ll 1$.
In this regime, the dipole interaction potential due to the standing wave $\mathbf{E}$ is given by
\begin{equation}
    V(z,t)=-\frac{1}{4}\text{Re}(\chi)|\mathbf{E}(z,t)|^2,
\end{equation}
where the polarizability $\chi$ is 
\begin{equation}
    \chi=4\pi\epsilon_0 R^3\frac{\epsilon(\omega)-1}{\epsilon(\omega)+2}=\epsilon_0\epsilon_c V.
\end{equation} 
In the latter expression the relative permittivity $\epsilon$ is the square of the complex refractive index $n(\omega)$ of the particle's material, $V=4\pi R^3/3$ is the volume of the particle, and we define $\epsilon_c=3(\epsilon-1)/(\epsilon+2)$. When the polarizable point-particle interacts with the standing wave grating, and ignoring  incoherent effects for the moment, its quantum state (reduced in the longitudinal direction) evolves unitarily as $\langle z|\psi\rangle\rightarrow\exp(i\phi_0 \cos^2 kz)\langle z|\psi\rangle,$ where the eikonal phase factor $\phi_0$ is obtained by integrating the dipole potential over the laser pulse duration~\cite{bateman2014near}
\begin{equation}\label{phiz}
    \phi(z)=\frac{1}{\hbar}\int_{\tau}dt V(z,t)=\phi_0 \cos^2 kz.
\end{equation}
In particular, we have that $\phi_0$ can be expressed in terms of the material polarizzability as well as the laser parameters as 
\begin{equation}
    \phi_0=\frac{2 \text{Re}(\chi) E_L}{\hbar c \epsilon_0 a_L},
\end{equation}
where $E_L$ is the pulse energy and $a_L$ is the spot area of the laser.

\subsection{Coherent grating for large particles}
Having briefly reviewed the point-particle case, we move now to consider spherical particles for which $kR\gtrsim 1$. We follow Ref.~\cite{nimmrichter2014macroscopic}, where the coherent effect of optical standing-wave gratings on extended spherical particles is analyzed. The expressions that we obtain in the following will allow us to construct the Talbot coefficients in the general case where incoherent effects are relevant. 

When the point-like particle approximation ceases to hold, a general treatment of light-matter interaction is in order since the particle can no longer be approximated by an electric dipole, and higher-order multi-poles should be considered. For homogeneous spherical particles, Mie scattering theory~\cite{mie1908beitrage,hulst1981light,bohren2008absorption} is appropriate. In fact, this theory offers exact solutions to Maxwell equations for light scattering from spherical objects. In order to derive the optical potential we look at the longitudinal light-induced force on the dielectric sphere. Note that, as remarked in Ref.~\cite{nimmrichter2014macroscopic}, transverse forces and corrections due to the finite-mode waist can be neglected owing to the short laser pulses that we consider. The light-induced forces acting on the dielectric particle can be obtained by integrating the electromagnetic stress-energy tensor over a spherical surface surrounding the particle. We follow Ref.~\cite{barton1989theoretical}, where a series-expansion expression of the net radiation force on a spherical particle of arbitrary size illuminated by a monochromatic light is obtained (see also Ref.~\cite{nimmrichter2014macroscopic}). It should be noted that, strictly speaking, Mie scattering theory considers plane electromagnetic waves. Thus, in obtaining the following expressions, the standing-wave profile of interest must be considered (see the Appendix~\ref{AppendixA}). We report here the expression for the longitudinal force on a dielectric sphere in vacuum in which we are interested,


\begin{widetext}
\begin{equation}
\label{shaub}
\begin{aligned}
 \frac{F_z(z)}{I_0 k^{-2}c^{-1}}&=-(kR)^4 \sum_{\ell=1}^{\infty} \sum _{m=\pm1} {\rm Im}\bigg[\ell(\ell+2)\sqrt{\frac{(\ell-m+1)(\ell+m+1)}{(2\ell+3)(2\ell+1)}}\\ 
 &\times\left.(2 a_{\ell+1,m}a_{\ell m}^*+ a_{\ell+1,m}A_{\ell m}^*+ A_{\ell+1,m}a_{\ell m}^*+2b_{\ell+1,m}b_{\ell m}^*+ b_{\ell+1,m}B_{\ell m}^* 
 +B_{\ell+1,m}b_{\ell m}^*)\right.\\
 &+m(2 a_{\ell,m}b_{\ell m}^*+a_{\ell,m}B_{\ell m}^*+A_{\ell,m}b_{\ell m}^*)\bigg],
\end{aligned}
\end{equation}
\end{widetext}
where $I_0=c\epsilon_0|E_0|^2/2$ is the intensity parameter of the incident light. This series contains several coefficients  -- $a_{\ell,m},b_{\ell,m},A_{\ell,m},B_{\ell,m}$-- that are derived starting from Mie scattering theory and that we report in the Appendix~\ref{AppendixA} for ease of exposition. 
As the longitudinal force due to the linearly polarized standing-wave $\mathbf{E}$ is of the form $F_{z}(z)=-F_0\sin 2kz$,  
the eikonal phase $\phi_0$ can be written as \footnote{We write the conservative force $F_{z}(z)$ as the gradient of a potential $V(z)$, i.e, $F(z)= -\nabla V(z)$. Inverting this relation we obtain $V(z)= -\frac{F_{0}}{k}\cos^2 (kz)$. Assuming a rectangular pulse of duration $\tau$, from Eq.~\eqref{phiz} we get $\phi_{0}= (F_{0}/k\hbar)\tau$. Finally, given the relation $\tau=8E_{L}/(c\epsilon_0 |E_{0}|^2 a_{L})$ between the impulse time and laser energy $E_{L}$ and spot area $a_{L}$, Eq.\eqref{phi0mie} is easily obtained.}
\begin{equation}\label{phi0mie} 
    \phi_0=\frac{8F_0 E_L}{\hbar c \epsilon_0 a_L k|E_0|^2}.
\end{equation}
In order to determine $F_0$, and thus $\phi_0$, we can just evaluate Eq.~\eqref{shaub} at $z=-\lambda/8$.
\begin{figure}[b!]
\centering
\includegraphics[width=\columnwidth]{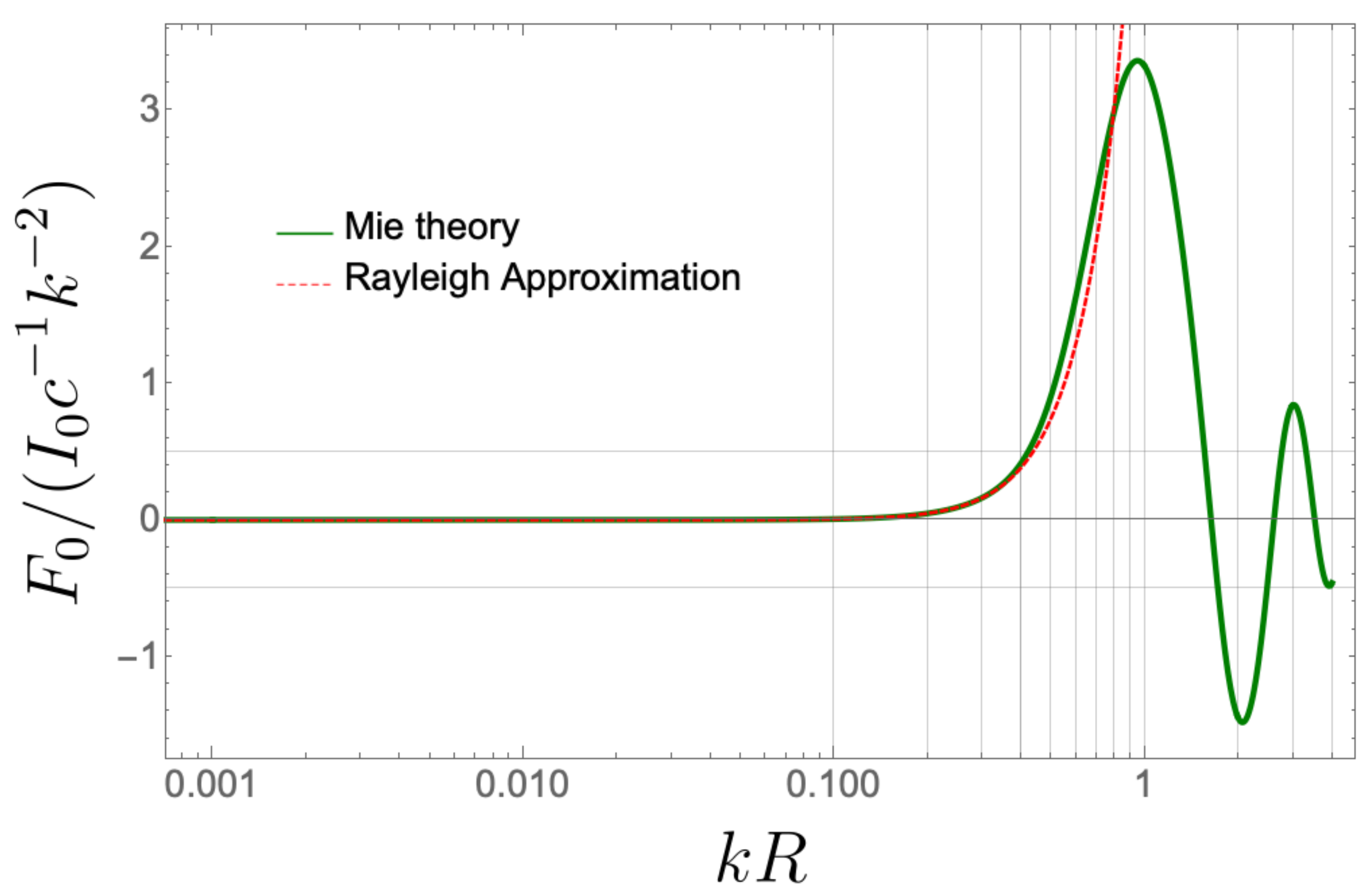}
  \caption{(Colors online) $F_0$ [in units of $I_0/(ck^2)$] as a function of the size parameter $kR$ for a silicon (Si) sphere at $\lambda=354$nm, where the \textit{bulk} refractive index~\cite{PhysRevB.27.985} is $n=5.656 + i\,2.952$. The solid green line is the result of Mie theory. The dashed red line is the prediction resulting from Rayleigh approximation.}
  \label{F0Si}
\end{figure}
Figure~\ref{F0Si} shows an example of its behaviour for a Si sphere at $\lambda=354$nm with the  {bulk} refractive index $n=5.656 + i 2.952$ (as tabulated in Ref.~\cite{PhysRevB.27.985}). As expected, the Rayleigh prediction (dashed red line) stops being valid for $kR\gtrsim 1$ and $F_0$ stops increasing with the volume of the sphere, showcasing an oscillatory behaviour. While a physical intuition behind this behaviour is provided in Ref.~\cite{nimmrichter2014macroscopic}, here we focus on the fact that, at the values of $kR$ corresponding to the nodes of the oscillations, the phase grating will be completely absent. This suggests that care should be used when choosing the size of large particles, so as to maximize the grating effects and avoid regions where the grating effect disappears. Moreover, one should bear in mind that the grating effect of the optical standing wave is greatly reduced for large particles with respect to the prediction of Rayleigh theory.

\begin{figure}[b!]
\centering
\includegraphics[width=\columnwidth]{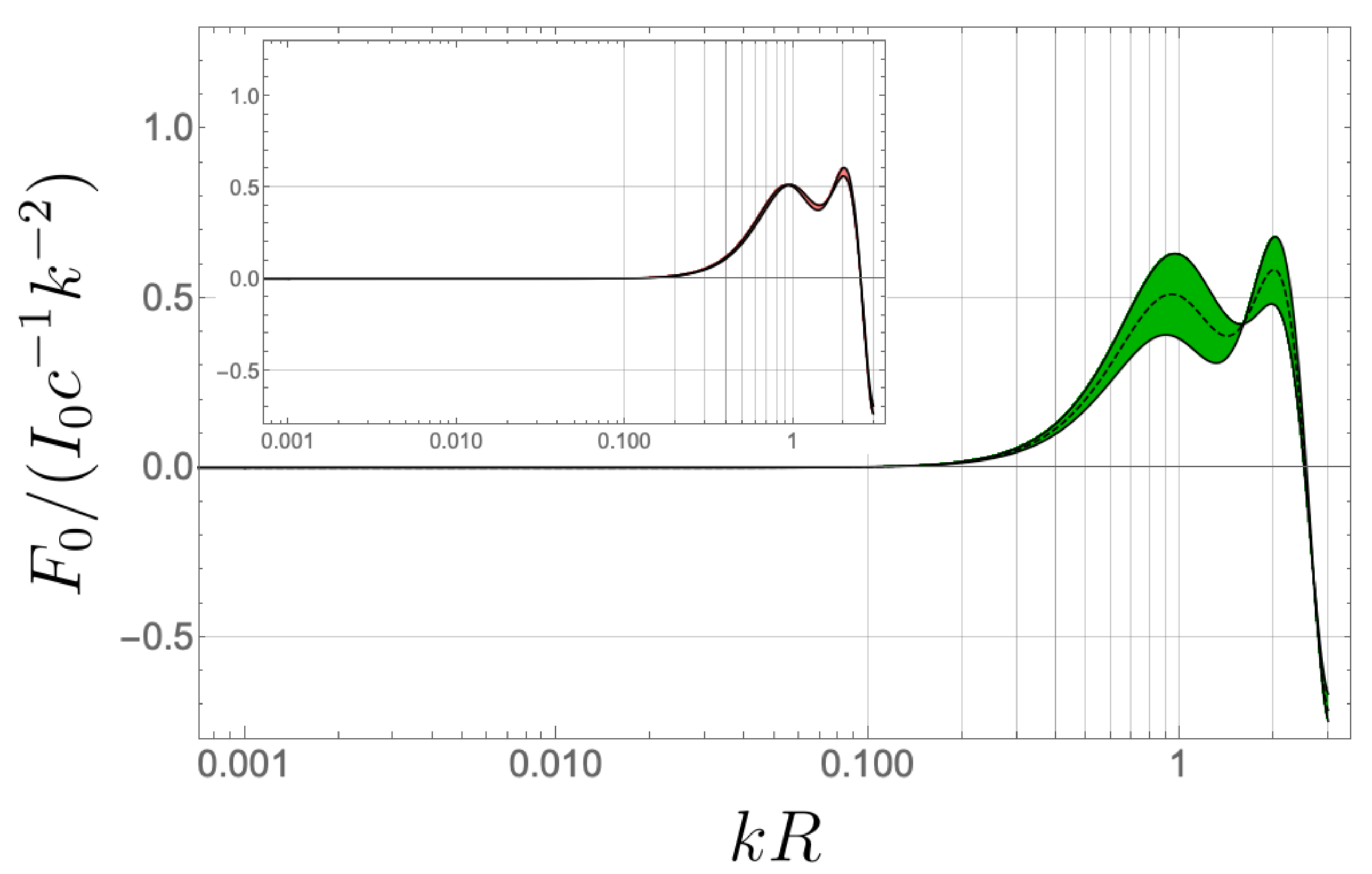}
  \caption{(Color online) $F_0$ [in units of $I_0/(ck^2)$] as a function of the size parameter $kR$ for a fused silica (SiO2) sphere at $\lambda=100$nm, where the \textit{bulk} refractive index has been roughly estimated as $n=1.3 + i\,0.8$ from tables in the Supplementary Material of~\cite{bateman2014near} (see also~\cite{kitamura2007optical}). The dashed black line is $F_0$ at $n=1.3 + i\,0.8$. The green region represents the result of a $\pm 5\%$ error in the real part of the refractive index. Inset: The red region represents the result of a $\pm 5\%$ error in the imaginary part of the refractive index. }
   \label{F0err}
\end{figure}

Another interesting point to consider here is the sensitivity of $F_0$ to changes in the refractive index. Indeed, it appears that the behaviour of $F_0$ against $kR$, can change significantly under variations of the refractive index. Figure~\ref{F0err} shows the effect of a $\pm 5\%$ variation in the value of the refractive index for fused silica at $100$nm. Fluctuations in the real part of $n$ can lead to quantitatively significant changes: the relative error in $F_0$ at the maxima can be as large as $10\%$ and even larger at the nodes. On the other hand, analogous inaccuracies on the imaginary part of $n$ lead to less important effects. 

We can conclude that, when doing Talbot-Lau interferometry with large spherical particles, the refractive index needs to be carefully estimated. Thus, the use of the bulk material refractive index could be too gross an approximation to the sphere refractive index\footnote{Note that, this effect may be very large for conducting nanoparticles, as the wave function of free electrons is sensitive to the size of the particle.}. This point deserves to be accounted for when planning experimental realizations.

\section{Incoherent effects}
\label{IV}

\subsection{Scattering}\label{IV.1}

In order to describe the incoherent effects due to scattering of standing wave photons we rely on the theory developed in~\cite{pflanzer2013optomechanics} for light-matter interaction in the Mie regime. According to~\cite{pflanzer2013optomechanics}, the effect of the scattering is described, under the assumption that the laser waist is much larger than the size of the particle, by the action of the Lindblad super-operator 
\begin{equation}
\label{eq:l_scatt}
\begin{aligned}
\mathcal{L}_{\rm{sca}}(\rho)& = |\alpha(t)|^{2}\int d\bold{k} \delta(\omega_{k}-\omega_{0})\\
&\times\left(2\mathcal{T}_{\bold{k,c}}(\hat{r})\rho\mathcal{T}^{*}_{\bold{k,c}}(\hat{r})-\left\{|\mathcal{T}_{\bold{k,c}}(\hat{r})|^{2},\rho \right\} \right),
\end{aligned}
\end{equation}
where the collisional operators are defined as
\begin{align}\label{eq:t_scatt}
\mathcal{T}_{\bold{k,c}}(\hat{r})= \frac{ic^{2}}{2\pi \omega}\int d\bold{k}'\, \braket{\bold{k}'|\bold{c}} f(\bold{k},\bold{k'})e^{-i(\bold{k}-\bold{k}')\cdot \hat{r}}
\end{align}
with $f(\bold{k},\bold{k'})$ the Mie scattering amplitude and $\ket{\bold{c}}$ the mode function of the standing wave, i.e., $\braket{r|\bold{c}}=1/\sqrt{V_{0}}\, f(x,y)\cos(kz)$ ($V_0$ is the mode volume of the standing wave).
Assuming that the free evolution is negligible during the interaction time, and working in the longitudinal eikonal approximation, the effect on the matter wave due to the scattering of grating photons is described by the action of the scattering mask $R_{\rm{sca}}(z,z')= e^{\int d\tau \mathcal{L}_{\rm{sca}}(z,z')}$ in the position representation, where

\begin{equation}
\label{annikaL}
\begin{aligned}
\mathcal{L}_{\rm{sca}}(z,z') &= |\alpha(t)|^{2}\int d\bold{k} \delta(\omega_{k}-\omega_{0})\left[2\mathcal{T}_{\bold{k,c}}(z)\mathcal{T}^{*}_{\bold{k,c}}(z')\right.\\
&\left.-|\mathcal{T}_{\bold{k,c}}(z)|^{2}-|\mathcal{T}_{\bold{k,c}}(z')|^{2} \right].
\end{aligned}
\end{equation}
In our case, the standing wave is  described in good approximation by $\mathbf{E}(\mathbf{r})\sim E_0 \mathbf{\hat{e}}_x \cos(kz)$. Thus, using the mode function $\braket{z|\bold{c}}=1/\sqrt{V_{0}}\cos(kz)$, one can show that 
\begin{align}\label{tsca}
\mathcal{T}_{\bold{k},\bold{c}}(z)&=  \sqrt{\frac{2\pi^3}{V_{0}}}\left(\mathcal{T}_{\bold{k_{0}},\bold{k}}^{*}(z)+\mathcal{T}_{-\bold{k_{0}},\bold{k}}^{*}(z)\right).
\end{align}
Substituting $z\rightarrow z_-=z-s/2 $ and $z'\rightarrow z_+=z+s/2$, and with the help of the above equation, we have
\begin{equation}
\label{eq:scatt_term}
\begin{aligned}
&\mathcal{L}_{\rm{sca}}\left(z_-,z_+\right) = \frac{|\alpha(t)|^{2}\pi c}{V_{0}}\left[\int d\Omega |f( k , k\bold{n})|^{2} \left(e^{-i (1 -n_{z}) ks}{-}1\right)\right.\\
&+ \int d\Omega f^{*}(k , k \bold{n})f(- k , k\bold{n})e^{-i2kz}\left(e^{ikn_{z}s}-\cos(ks)\right)\\
&+\int d\Omega f^{*}(-k , k \bold{n})f(k , k\bold{n})e^{i2kz}\left(e^{ikn_{z}s}-\cos(ks)\right)\\
&\left.+ \int d\Omega |f(-k , k\bold{n})|^{2}\left(e^{i(1+n_{z})ks}-1\right)\right],
\end{aligned}
\end{equation}
where $\mathbf{n}= \bold{k'}/|k|,$ and $\Omega$ is the solid angle associated with $k'$ assuming that $\bold{k}$ is pointing in the $z$-direction.  

We note that the spherical symmetry of the nano-particle is reflected in the following symmetry of the scattering amplitude\footnote{Note that, in the Rayleigh approximation a further symmetry appears because the particle is treated as point-like. In particular, $f(-k,k\bold{n})= f(k,k\bold{n})$. Employing this symmetry, it is straightforward to obtain the Rayleigh scattering expressions from Eq.~\eqref{eq:scatt_term}~\cite{nimmrichter2014macroscopic}.} $f(-k,k\bold{n})= f(k,-k\bold{n})$.
Exploiting the symmetry, and through lengthy but otherwise straightforward algebra, we finally get 
\begin{equation}\label{kern-sca}
R_{\rm{sca}}\left(z_-,z_+\right)= \exp(F(s)+a(s)\cos(2kz)+ib(s)\sin(2kz))
\end{equation}
where
\begin{widetext}
\begin{align}\label{abF}
a(s) &= \frac{2\pi c}{V_0}\int d\tau|\alpha(\tau)|^{2}\int d \Omega\,{\rm Re}\big(f^{*}(k , k \bold{n})f(- k , k\bold{n})\big)[\cos(k n_{z} s)-\cos(ks)],\nonumber\\
b(s)&=  \frac{2\pi c}{V_0}\int d\tau|\alpha(\tau)|^{2}\int d \Omega \,{\rm Im} \big(f^{*}(k , k \bold{n})f(- k , k\bold{n})\big)\sin(kn_{z}s),
\nonumber\\
F(s)&=  \frac{2\pi c}{V_0}\int d\tau |\alpha(\tau)|^{2}\int d \Omega\, |f(k,k\bold{n})|^{2} [\cos((1-n_{z})ks)-1].
\end{align}
\end{widetext}
Henceforth, for ease of notation, we omit the explicit dependence on $s$ of the functions $a(s)$, $b(s)$,  and $F(s)$.
Note that $\int d\tau |\alpha(\tau)|^2$ can be expressed in terms of the laser pulse parameters as $\int d\tau |\alpha(\tau)|^2=4V_0 E_L/(\hbar c\omega\,a_L)$.
We can now compute the Fourier coefficients of the scattering mask $R_{\rm{sca}}(z,z')$ which enters in the Talbot coefficients of Eq.~\eqref{eq:TalbotCoeff} 
\begin{equation}\label{FourSca}
\begin{aligned}
R_{n}\left(\frac{s}{d}\right)&= \frac{1}{d} \int_{-d/2}^{d/2} dz e^{i 2\pi \frac{n z}{d}}e^{a\cos(\frac{2\pi z}{d})+ib\sin(\frac{2\pi z}{d})+F}\\
&= e^{F}\left(\frac{a-b}{a+b}  \right)^{n/2} \text{I}_{n}(\sign(a-b)\sqrt{a^{2}-b^{2}}),
\end{aligned}
\end{equation}
where $I_{n}(a)$ are modified Bessel functions of the first kind.
We can then use Graf's addition theorem~\cite{erdelyi1964m} to rewrite the Fourier coefficients as 
\begin{align}
R_{n}\left(\frac{s}{d}\right)=e^{F}\sum_{k}I_{k+n}(a)J_{n}(b),
\end{align}
where $J_{n}(b)$ are Bessel functions of the first kind.
Exploiting this result we can conveniently write the Talbot coefficient, modified by the presence of scattering mechanisms, as
\begin{equation}
\tilde{B}_{n}\left(\frac{s}{d}\right)= e^{F}\sum_{k,m} J_{n-k}(\xi_{coh}) I_{k+m}(a)I_{k}(b),\\
\end{equation}
where $\xi_{coh} = \phi_{0}\sin\left(\frac{\pi s}{d}\right)$, and using again Graf's theorem obtain
\begin{equation}
\tilde{B}\left(\frac{s}{d}\right) =\sum_{k }\Lambda^{\frac{n+k}{2}}J_{k+n}\left(\sign(\zeta_{coh}-a)\sqrt{\zeta_{coh}^{2}-a^{2}} \right)J_{n}(b)
\end{equation}
with $\Lambda=e^{F}(\zeta_{coh}+a)/(\zeta_{coh}-a)$.
\subsection{Absorption}\label{IV.2}
Apart from the scattering of grating photons, also their absorption gives rise to an additional incoherent effect. This effect is relevant unless very low-absorbing material spheres are employed, and it is amplified by the size of the spheres. However, while a quantum formalism beyond the point-particle approximation exists for the description of scattering of grating photons, no such formalism is available for absorption. 

In order to estimate the incoherent effect due to absorption, we follow here a semiclassical approach, fostered in Ref.~\cite{nimmrichter2014macroscopic}, which is valid in the Rayleigh regime. Nonetheless, we will improve on it by considering the actual number of absorbed photons, which depends on the Mie absorption cross-section. The result coming from this analysis only embodies a rough estimate of the real effect of absorbed photons and, in general, results in a lower bound on the actual amount of decoherence. We comment on how to possibly extend this approach at the end of this Section.

In Ref.~\cite{nimmrichter2014macroscopic} a Lindblad (super)-operator describing the incoherent effect of photon absorption is obtained by treating absorption as a Poisson process and using the corresponding noise in a stochastic Schr\"odinger equation describing the evolution of the state of an absorbing particle. The end result, after averaging, is a Lindblad-like equation with jump operators describing the evolution of the state of the particle every time an absorption event occurs. In order to determine the jump operator characterizing absorption, consider a spherical particle in the Rayleigh limit with complex polarizability, and the effect of absorbing a photon from the linearly polarized incident light with the  mode function $f(\mathbf{r})\hat{e}_x$. As the mode function can be expanded on the basis of plane waves $f(\mathbf{r})=\sum f_\mathbf{k} \exp{i \mathbf{k}\cdot\mathbf{r}}$, and the absorption of a plane wave photon amounts to a shift in momentum space by $\hbar\mathbf{k}$, the effect of absorbing a photon from the incident light transforms a momentum state of the particle as
\begin{equation}
    |\mathbf{p}\rangle\rightarrow \sum f_\mathbf{k} |\mathbf{p}+\hbar\mathbf{k}\rangle=f(\mathbf{r}) |\mathbf{p}\rangle.
\end{equation}
Thus, the effect of the jump operator is given simply by the scalar mode function. The rate at which the absorption occurs is related to the number of photons in the light field $|\alpha|^2$ times the single photon absorption rate $c\sigma_{\rm{abs}}/V_0$, where $\sigma_{\rm{abs}}$ is the absorption cross-section and $V_0$ is the mode volume of the incident light. 

For a standing wave, and neglecting the effect on the transverse motion of the particle, the action of the Lindblad operator on the particle's density matrix reads 
\begin{equation}\label{abs}
    \mathcal{L}_{\rm{abs}}(\rho)=\frac{c\sigma_{\rm{abs}}}{V_0}|\alpha(t)|^2\left[\cos(kz)\rho\cos(kz)-\frac{1}{2}\{ \cos^2(kz),\rho\}\right],
\end{equation}
where the time dependence of $|\alpha|^2$ reflects the fact that we are considering a pulse.

In order to better estimate the effects of absorption for large particles, a first crude approximation is to consider the same Lindblad super-operator as in Eq.~\eqref{abs} whilst considering the right absorption cross section as given by Mie scattering theory. This is given by $\sigma _{\text{abs}}=\sigma _{\text{ext}}-\sigma _{\text{sca}}$, i.e. the difference between the total extinction cross-section and the scattering one. Explicitly
\begin{equation}
    \sigma _{\text{abs}}=\frac{2 \pi(2 n+1)  }{k^2}\sum_{n=1}^{\infty }\left({\rm Re}\left(a_n+b_n\right)-|a_n|^2-\left|b_n|^2\right.\right),
\end{equation}
in terms of the Mie coefficients, which are given in the Appendix~\ref{AppendixA}.

With these expressions at hand, we can again follow the steps in Sec.~\ref{IV.1}, while including also the absorption super-operator. The decoherence mask due to absorption is given by
\begin{align}\label{kern-abs}
    R_{\rm{abs}}(z,z')=\exp\left[-2n_0 \sin^2\left(k_0 \frac{z+z'}{2}\right)\sin^2\left(k_0 \frac{z-z'}{2}\right)\right],
\end{align}
where $n_0$ is the mean number of absorbed photons at the anti-nodes
 $ n_0=\frac{4\sigma_{abs}}{h c}\frac{E_L}{a_L}\lambda=\frac{I_0}{c F_0}\sigma_{abs} \phi_0$.
Note that, by substituting $z\rightarrow z_- $ and $z'\rightarrow z_+$ and including also the effect of scattering, Eq.~\eqref{FourSca} is modified to
\begin{widetext}
\begin{equation}
\begin{aligned}
    R_{n}\left(\frac{s}{d}\right)&=\frac{e^{F}}{d}\int_{-d/2}^{d/2}dz\, e^{-2\pi i n \frac{z}{d}}e^{a\cos\left(\frac{2\pi z}{d}\right)-b\sin\left(\frac{2\pi z}{d}\right)-2n_0 \sin^2\left(\pi \frac{z}{d}\right)\sin^2\left(\frac{\pi s}{2d}\right)} \\
&=\frac{e^{F-\frac{c_{\rm{abs}}}{2}}}{d}\int_{-d/2}^{d/2}dz e^{-2\pi i n \frac{z}{d}}e^{(a+\frac{c_{\rm{abs}}}{2})\cos\left(\frac{2\pi z}{d}\right)-b\sin\left(\frac{2\pi z}{d}\right)},
\end{aligned}
\end{equation}
\end{widetext}
where $c_{\rm{abs}}=n_0 (1-\cos(\pi s/d))$.  
We can now follow exactly the same steps as for the scattering case and end up with new Talbot coefficients that include absorption
\begin{equation}
\begin{aligned}
&\tilde{B}_{n}\left(\frac{s}{d}\right)= e^{F-c_{\rm{abs}}/2} \sum_{k=-\infty}^{\infty}\left(\frac{\zeta_{\rm{coh}}+a+c_{\rm{abs}}/2}{\zeta_{\rm{coh}}-a-c_{\rm{abs}}/2}\right)^{\frac{n+k}{2}}J_{k}(b)\\
&\times J_{n+k}\left(\sign(\zeta_{\rm{coh}}-a-c_{\rm{abs}}/2)\sqrt{\zeta_{\rm{coh}}^{2}-(a+c_{\rm{abs}}/2)^{2}}\right).
\end{aligned}
\end{equation}
It should be noted that, the only difference between the expressions for absorption presented here and the ones in, e.g.,~\cite{nimmrichter2014macroscopic} is in the use of the Mie absorption cross-section. 

The treatment of the absorption decoherence is based on a semiclassical approach in the Rayleigh limit, i.e., treating the particle as point-like. While we have refined the result by using the Mie theory absorption cross section, the formalism does not properly account for the finite size of the particle and the variation of the light intensity across it. In order to extend the formalism beyond the Rayleigh approximation, we look at the non-conservative part of the classical force acting on a polarizable particle of finite size interacting with the electromagnetic field
\begin{align}
F_{\rm{nc}}(\bold{r})= -\frac{\epsilon_{0} \epsilon_{c}^{\scriptscriptstyle{I}}}{2} \int_{V_{n}(r_{t})} d\bold{r}\, {\rm Im}
\{[\nabla \cdot E^{*}(\bold{r})] E(\bold{r})\}, 
\end{align}
where $\epsilon_{c}^{\scriptscriptstyle{I}}$ is the imaginary part of $\epsilon_{c}$. The appearance of a non-conservative force is an artifact of having ignored the dynamics of the internal degrees of freedom (d.o.f.s) of the particle. If the latter were to be included, the complete dynamics would be fully unitary and no non-conservative force would appear. While, as far as we know, a full model for dielectric particles is not present, for a single atom interacting with a single quantized field mode such a treatment is viable~\cite{jaynes1963comparison,boukobza2005entropy} and indeed leads to the absorption and scattering of photons by the atom. Notwithstanding the technical details, from the form of the non-conservative classical force we could argue that, replacing $\nabla\cdot E$ with the particle charge density $\rho_q$ and including its dynamics, a potential term coupling the incident field with the internal phonons modes would arise. These terms will be analogous to the coupling between the incident and the scattered electric field used in~\cite{pflanzer2012master}, to derive Eq.~\eqref{annikaL}, with now, instead of the scattering amplitudes, the absorption ones.
This suggests that the right form of the absorption term should be similar to Eq.~\eqref{eq:scatt_term} with the appropriate amplitudes and phononic mode functions. However, as already mentioned, a microscopic model for the interaction between the internal d.o.f.s and light is currently missing.

\subsection{Classical Limit}\label{ClaLim}
In near-field matter interferometry, a fringe pattern may also appear when a classical description of the particle -- in terms of ballistic trajectories --  is adopted. This is due to partial reflection by the light grating~\cite{hornberger2009theory}. Therefore, a non-vanishing fringe contrast is not sufficient to prove genuine quantum interference and one would have to resort to a direct comparison between the quantum and the classical models for the dynamics. The classical behaviour can be obtained as the limiting case of Eqs.~\eqref{eq:conv_kern_tr} and~\eqref{eq:conv_kern_deco} for $\hbar \to 0$ (applied to a classical probability distribution in phase space). Here, we do so for the coherent and incoherent convolution kernels, for both the scattering and the absorption case, as the $\hbar\rightarrow 0$ limit of the quantum expression. 

\begin{figure*}[t!]
\centering
\includegraphics[width=\textwidth]{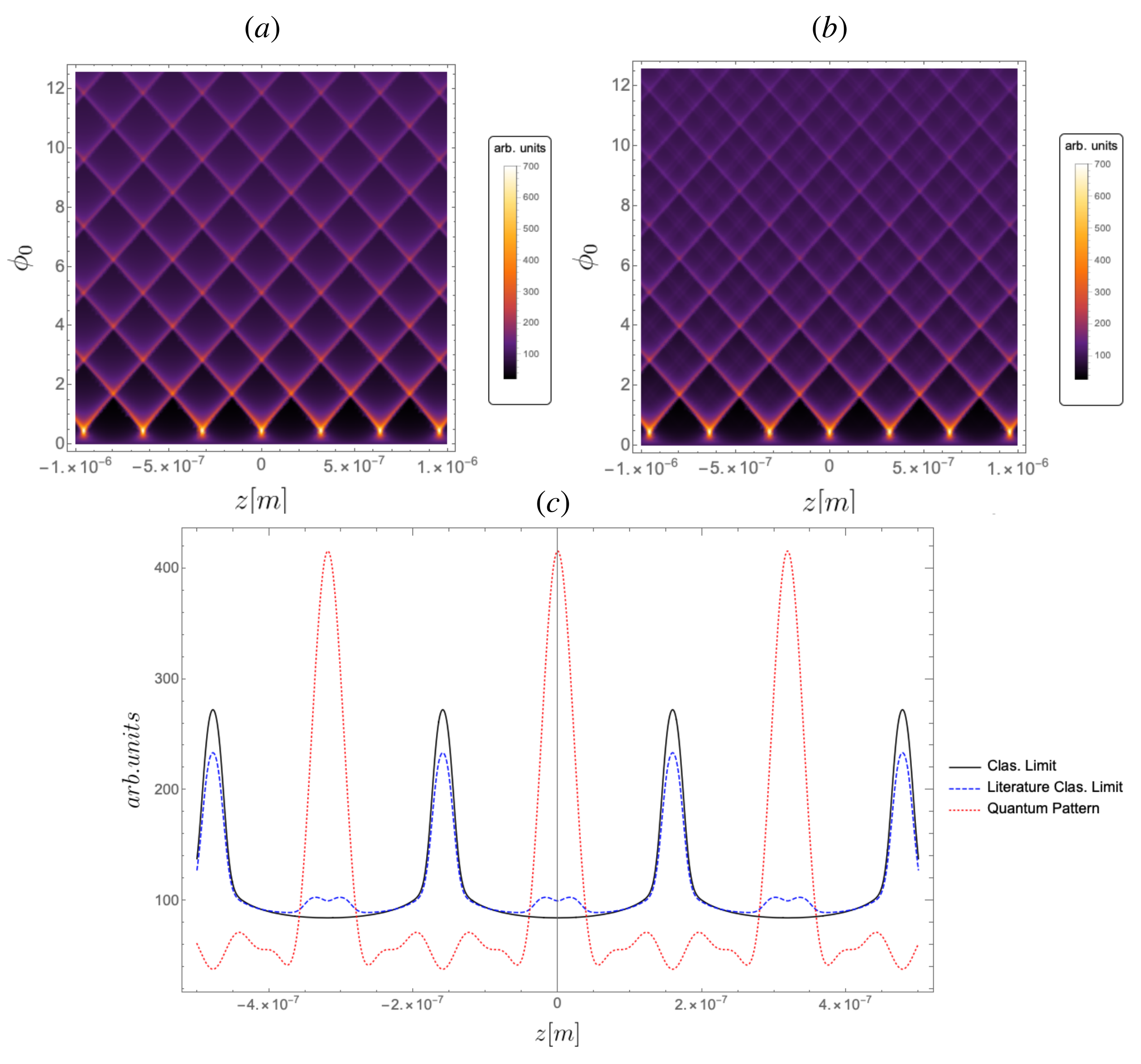}
  \caption{(Color online) Effect of the optical grating, photon scattering, and absorption on the interference pattern in the classical limit. Here we use the parameters in Table~\ref{tab} with $m=10^6$ amu and the corresponding $x=kR\sim 0.098$, where the use of the Rayleigh limit is well justified. We have assumed the same experimental arrangements and configuration as in Ref.~\cite{bateman2014near}, with the exception of (gravitational) acceleration and environmental decoherence effects during the free evolution times $t_1, t_2$, which have been neglected to highlight the effects of the grating, scattering and absorption mechanisms. (a) The classical fringe pattern, when varying the maximum phase modulation $\phi_0$, without decoherence, i.e., what we obtain by taking the classical limit as discussed in the text. (b) The decoherence terms due to scattering and absorption are evaluated using the same integral kernels as for the quantum case. This is what has been advocated previously in the literature (cf. Ref.~\cite{PhysRevA.70.053608}). (c) An instance of the interference patterns for $\phi_0=4$. The solid black curve corresponds to panel (a); the dashed blue curve to (b); and the dotted red curve, to the quantum interference pattern including decoherence due to scattering and absorption of the grating photons. Due to different notations, the Rayleigh limit of the Lindblad super-operator in
  Eq.~\eqref{eq:scatt_term} differs from Eq.~(2.24) in Ref.~\cite{nimmrichter2014macroscopic} by a factor of $1/4\pi$.}
   \label{ClaLimfig}
\end{figure*}

To show how the limit is performed, it is convenient to consider first the coherent convolution kernel Eq.~\eqref{eq:conv_kern_tr}
\begin{align}\label{eq:conv_kern_tr2}
\mathcal{T}_{\rm{coh}}(z,p) &= \frac{1}{2\pi \hbar}\int ds\,e^{\frac{ips}{\hbar}}e^{-\frac{i}{\hbar}\int_{0}^{t}d\tau [V(z-s/2,\tau)-V(z+s/2,\tau)]},
\end{align}
where we have used Eq.~\eqref{Tco}. We first rescale the integration variable as $s\rightarrow s\hbar$ to have $\hbar$ appearing only in the argument of the exponential, and then Taylor expand the potential $V(z\pm s\hbar)$ to first order in $\hbar$. In this way, we get 
\begin{align}\label{eq:tclass}
\mathcal{T}_{\rm{coh}}(z,p) &\simeq\frac{1}{2\pi}\int ds\,e^{i\,s\,\left(p+\int_{0}^{t}d\tau \nabla V(z,\tau)\right)+\mathcal{O}(\hbar)},
\end{align}
which, upon taking the limit $\hbar\to0$, gives us the classical convolution kernel 
\begin{equation}
    \mathcal{T}_{\rm{class}}(z,p)=\delta\left(p + \int d\tau \nabla V(z,\tau)\right).
\end{equation}
Following the same logic, the classical limit of the decoherence convolution kernels can be obtained. It should be noted that, re-scaling $s$ and then expanding around $\hbar=0$ is equivalent to only performing an expansion around $s=0$ of the argument of the exponential in the Fourier transform of the kernels.  

\begin{figure*}[t!]
\centering
\includegraphics[width=\textwidth]{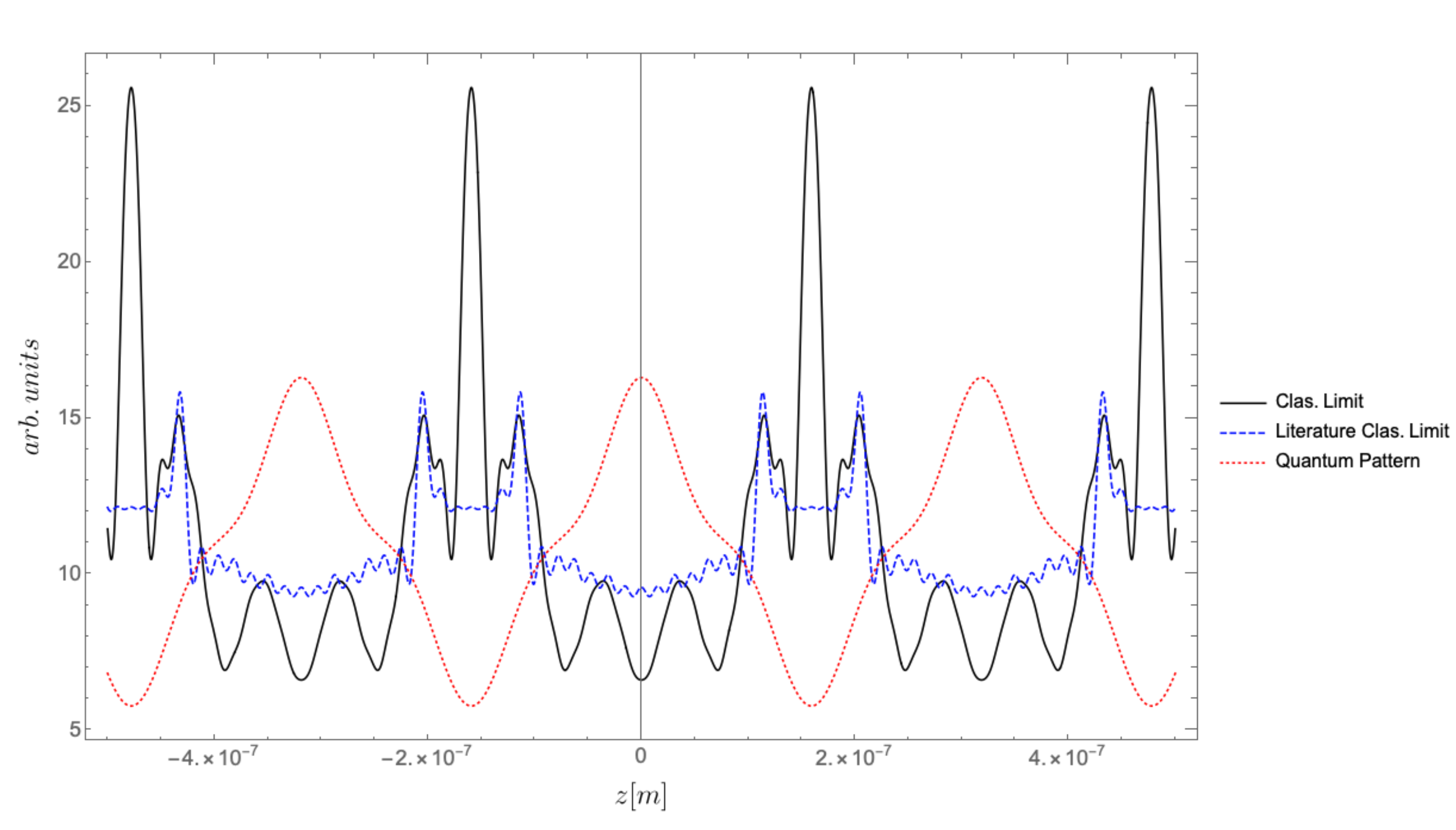}
  \caption{(Color online) Effect of the coherent grating and photon scattering on the interference pattern in the classical limit. Here we used the parameter in Table~\ref{tab} with a mass $m=10^8$amu, which corresponds to a value of $x=kR\sim 0.46$, for which the Rayleigh approximation is not well-justified.
  The figure shows an instance of the interference patters for $\phi_0=2$, where only the decoherence due to scattering of grating photons, of which we have full control, is considered. 
   The solid black curve corresponds to the classical limit as obtained in this work. The dashed blue curve corresponds to the classical limit obtained without modifying the incoherent kernels from the quantum case. The dotted red curve represents the quantum case. Note the striking difference between the two classical interference patters, which is larger with respect to the Rayleigh case given the greater effect of scattering decoherence in the Mie regime.}
   \label{ClaLim2}
\end{figure*}

Consider Eq.~\eqref{eq:conv_kern_deco} with $R_{{\rm sca}}(z_-,z_+)$ given by Eq.~\eqref{kern-sca}. Rescaling the integration variable, the dependence on $\hbar$ occurs only in the functions $a,\,b,\, F$ in Eq.~\eqref{abF}. It is crucial to note at this stage that, for the electromagnetic field, the identification of $|\alpha|^2$ with the classical intensity $I(\tau)$ over $\hbar$ requires expanding the trigonometric functions in $a,b, F$ to first order in $\hbar$, in analogy with the coherent kernel. It is easy to see that the only non-vanishing contribution to the classical decoherence kernel arises from the function $b$. The same argument shows that, the classical limit of the absorption decoherence kernel in Eq.~\eqref{kern-abs} vanishes, making no contribution to the classical dynamics. However, a more refined treatment of absorption decoherence --- along the lines depicted in Sec.~\ref{IV.2} --- should make a contribution similar to the one found for scattering. It is interesting to note that, treating the light-matter interaction in the Rayleigh limit would lead to the complete absence of a decoherent effect for the classical dynamics\footnote{In the Rayleigh approximation, the $b$ term in Eq.~\eqref{abF}, is identically zero due to the symmetry property of the scattering amplitude.}, in contrast to what has been argued in the existing literature. This observation could already be relevant for upcoming experiments working in the Rayleigh approximation, as it would help identify the working point at which the differences between classical and quantum interference patterns are maximum (cf. Figs.~\ref{ClaLimfig} and~\ref{ClaLim2}).

\section{Realistic example}\label{V}
We now study the effects of scattering and absorption of grating photons on the interference pattern. 
The figure of merit embodied by the sinusoidal fringe visibility $\mathcal{V}_{\rm{sin}}$ is used here to complement the predictions coming from  the interference pattern. As outlined in Refs.~\cite{nimmrichter2014macroscopic,bateman2014near}, the interference pattern is often dominated by the first Fourier amplitude. Thus, the fringe contrast 
is well described by (cf. the caption of Table~\ref{tab})
\begin{equation}
    \mathcal{V}_{\rm{sin}}=2\Bigg|\tilde{B}_{1}\Bigg(\frac{t_1 t_2}{t_T (t_1+t_2)}\Bigg)\Bigg|\exp\Bigg(-\frac{2\pi^2\sigma_z^2t_2^2}{d^2(t_1+t_2)^2}\Bigg).
\end{equation}
Here we focus on the fringe visibility of the quantum interference pattern. We follow a recent proposal for an experimental realization of matter-wave interferometry with nano-particles~\cite{bateman2014near} from which the parameters in Table~\ref{tab} have been drawn. There, silicon (Si) nano-particles with a mass of $10^6$amu are considered. For such a mass, the scattering of grating photons is completely negligible, while the effect of absorption is not insignificant. Nevertheless, the results obtained from Mie and the Rayleigh theory for both scattering and absorption are in good agreement, as it should be expected given the value $x=k R\sim 0.098$ of the parameter controlling the validity of the point-like approximation. 
\begin{table}[t]
\centering
\begin{tabular}{|l|l|l|l|l|}
\hline
\multicolumn{5}{|l|}{Laser: $\lambda=2d=354\times 10^{-9} \rm{m}$}                                         \\ \hline
$\rho_{\rm{Si}}=2.3290 \times10^3 \rm{Kg}/\rm{m}^3$ & \multicolumn{4}{l|}{$T=20 \times 10^{-3}$K}          \\ \hline
\multicolumn{5}{|l|}{Refractive Index at $\lambda$: $n=5.656 + i\,2.952$}                                  \\ \hline
\multicolumn{5}{|l|}{Trapping frequency: $\nu=200\times 10^3$Hz}                                           \\ \hline
\multicolumn{5}{|l|}{Interferometer:}                                                                      \\ \hline
$d=177 \times 10^{-9} \rm{m}$                       & $t_1=2t_T$       & \multicolumn{3}{l|}{$t_2=1.6 t_T$} \\ \hline
\end{tabular}
\caption{Parameters considered for Si spheres. Other parameters entering the generalized Talbot coefficients and the interference pattern (cf. Eq.~(4) in Ref.~\cite{bateman2014near}) can be inferred from the table. In particular, $\sigma_z=\sqrt{k_B T/(4 \pi ^2 m \nu^2)}$, $D=d (t_1+t_2)/t_1$, and the Talbot time $t_T=d^2 m/h$. The mass $m$ of the spheres enters also into the definition of the sphere radius via $R^3=(3/4\pi)(m/\rho_{Si})$; the greater the mass, the larger its radius.}
\label{tab}
\end{table}

\begin{figure*}[t!]
\centering
\includegraphics[width=\textwidth]{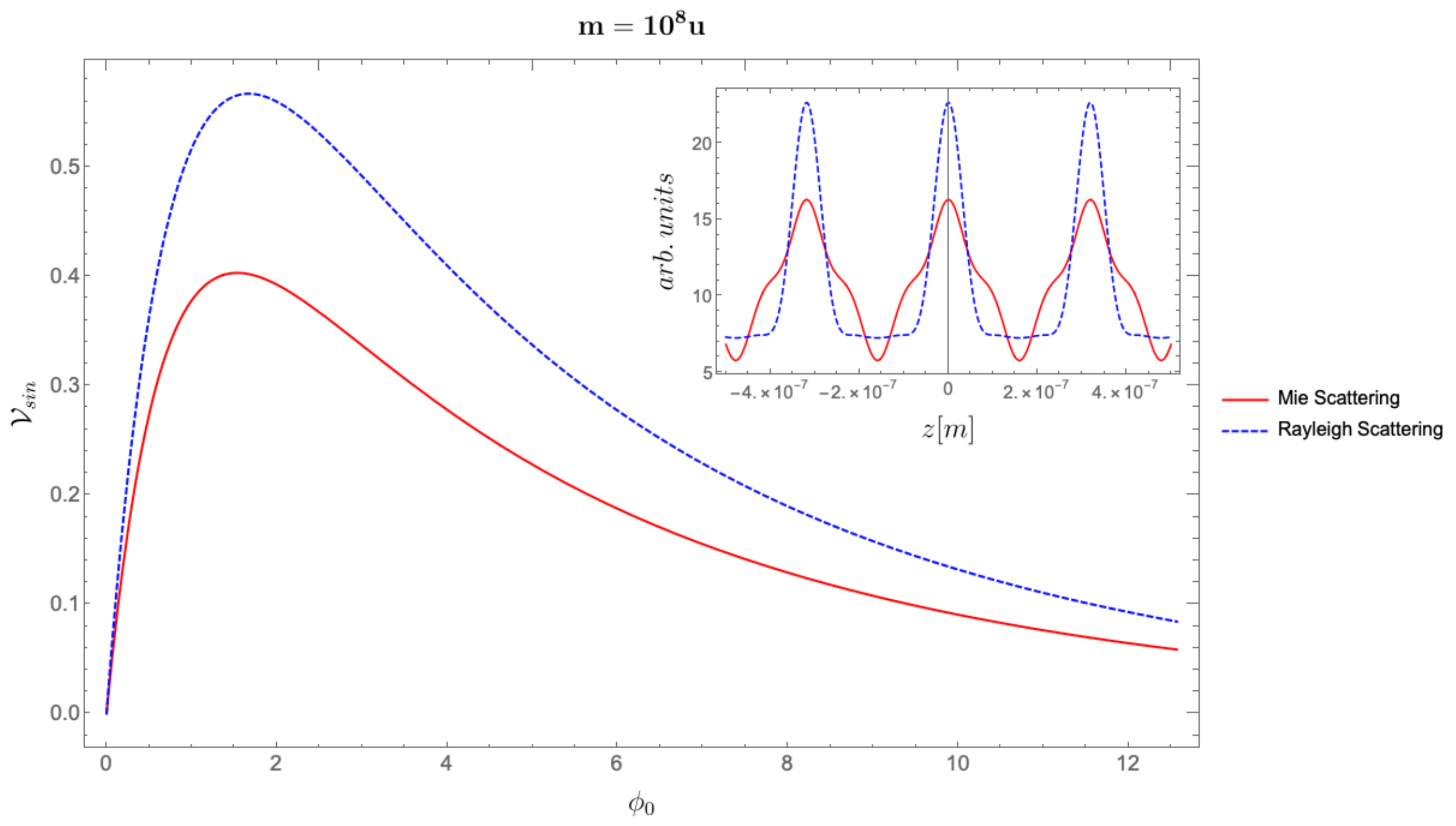}
  \caption{(Color online) Sinusoidal visibility of the interference pattern for a particle of mass $m=10^8$amu, which corresponds to a value $x=kR\sim 0.46$ for the parameters reported in Table~\ref{tab}. For these parameters, the use of the Rayleigh approximation is not fully justified. The dashed  blue curve represents the interference pattern visibility in the case in which Rayleigh approximation is used. The solid red curve represents the visibility when Mie theory is employed. Inset: An instance of the interference patters for $\phi_0=2$ showing a significant difference between the two situations.}
   \label{FigMie1}
\end{figure*}

However, a mass of $10^8$amu makes $x=kR\sim 0.46$. Although this is a modest increase with respect to the previous case, it turns out that the Rayleigh approximation is no longer well justified. Figure~\ref{FigMie1} shows the difference in the predictions obtained 
 using the Rayleigh approximation versus those arising from Mie theory. It should be noted that, in Fig.~\ref{FigMie1} we consider only decoherence effects due to scattering of grating photons, of which we have full control. Thus, no decoherence effect due to black-body radiation, gas particle collisions, or, most importantly, absorption of grating photons is considered in this case. As can be easily seen, the visibility is strongly affected. Even more significantly, the form of the interference pattern is significantly modified, the deformation being even more important at higher values of the mass parameter.
 
Finally, we note that, while we employ the sinusoidal visibility to show the results associated with the use of Mie theory, this is not a very useful indicator when it comes to comparing the quantum prediction for the interference pattern with the classical shadow pattern. Indeed, while the quantum visibility could be smaller than its counterpart corresponding to the classical pattern, the interference figures could still be clearly distinguishable due to the position and shape of the oscillatory peaks. We stress here that better figure of merit should be used in order to certify the quantumness of an observed interference pattern.

\section{Discussion and Conclusion}\label{VI}

We have addressed the Talbot-Lau effect beyond the Rayleigh limit, accounting for the suppression of coherent grating effects due to large-size particles, scattering and photon absorption.  
 
 We have only considered polarizable spherical particles and neglected their internal degrees of freedom. These approximations are the usual workhorse in many matter-wave experiments and allow us to neglect decoherence effects due to coupling between the center-of-mass motion and other degrees of freedom.   

The main results of this work are the expressions needed to describe the coherent and incoherent effects beyond the Rayleigh approximation due to optical grating. Nonetheless, the discussion of the classical limit provides some interesting insight. Indeed, we have shown that the classical limit of the decoherent effects due to scattering and absorption of grating photons is qualitatively different from the results presented in the current literature. In particular, it appears that when the Rayleigh approximation is well justified, then, in the classical limit, no effect due to scattering and absorption survives, leaving only the deflection of ballistic trajectories due to the standing wave. While relevant for current experimental proposals, this result has striking consequences also for future experiments aiming to study large particles superpositions as it significantly reduces the decoherence suppressing the would-be classical shadow effect from which the quantum interference pattern needs to be distinguishable.

Our study is motivated by the need to account for  increasing sizes of particles in experiments aiming at probing the quantum-to-classical transition. However, a number of assumptions were necessary in order to develop our framework. Two of them are particularly relevant for future endeavours: spherical symmetry of the particles, and their homogeneity. While it could still be a very good approximation, the Mie theory is not rigorously applicable when such assumptions are relaxed. Moreover, additional decoherence effects can arise due to the coupling between the center of mass and the rotational degrees of freedom of a non-spherical and/or an-isotropic object, and by coupling with the internal degrees of freedom. The assessment of such questions will be the focus of our future investigations. 

\section{Acknowledgements}
We would like to thank J. Bateman and S. Nimmrichter for useful discussions and suggestions. A.B. is grateful to Eric Lutz for the hospitality of his group at Universit\"at Stuttgart during completion of this work. The authors gratefully acknowledge support from the MSCA IEF project Perfecto (grant nr. 795782), the EU Collaborative project TEQ (grant nr. 766900), the DfE-SFI Investigator Programme (grant 15/IA/2864), COST Action CA15220, the Royal Society Wolfson Research Fellowship ((RSWF\textbackslash R3\textbackslash183013), the Leverhulme Trust Research Project Grant (grant nr.~RGP-2018-266). R.K. acknowledges support from the Austrian Research Promotion Agency (FFG, Projects Nos. 854036 and 865996).

\bibliography{references2.bib}

\begin{thebibliography}{31}%
\makeatletter
\providecommand \@ifxundefined [1]{%
 \@ifx{#1\undefined}
}%
\providecommand \@ifnum [1]{%
 \ifnum #1\expandafter \@firstoftwo
 \else \expandafter \@secondoftwo
 \fi
}%
\providecommand \@ifx [1]{%
 \ifx #1\expandafter \@firstoftwo
 \else \expandafter \@secondoftwo
 \fi
}%
\providecommand \natexlab [1]{#1}%
\providecommand \enquote  [1]{``#1''}%
\providecommand \bibnamefont  [1]{#1}%
\providecommand \bibfnamefont [1]{#1}%
\providecommand \citenamefont [1]{#1}%
\providecommand \href@noop [0]{\@secondoftwo}%
\providecommand \href [0]{\begingroup \@sanitize@url \@href}%
\providecommand \@href[1]{\@@startlink{#1}\@@href}%
\providecommand \@@href[1]{\endgroup#1\@@endlink}%
\providecommand \@sanitize@url [0]{\catcode `\\12\catcode `\$12\catcode
  `\&12\catcode `\#12\catcode `\^12\catcode `\_12\catcode `\%12\relax}%
\providecommand \@@startlink[1]{}%
\providecommand \@@endlink[0]{}%
\providecommand \url  [0]{\begingroup\@sanitize@url \@url }%
\providecommand \@url [1]{\endgroup\@href {#1}{\urlprefix }}%
\providecommand \urlprefix  [0]{URL }%
\providecommand \Eprint [0]{\href }%
\providecommand \doibase [0]{http://dx.doi.org/}%
\providecommand \selectlanguage [0]{\@gobble}%
\providecommand \bibinfo  [0]{\@secondoftwo}%
\providecommand \bibfield  [0]{\@secondoftwo}%
\providecommand \translation [1]{[#1]}%
\providecommand \BibitemOpen [0]{}%
\providecommand \bibitemStop [0]{}%
\providecommand \bibitemNoStop [0]{.\EOS\space}%
\providecommand \EOS [0]{\spacefactor3000\relax}%
\providecommand \BibitemShut  [1]{\csname bibitem#1\endcsname}%
\let\auto@bib@innerbib\@empty
\bibitem [{\citenamefont {Hornberger}\ \emph {et~al.}(2012)\citenamefont
  {Hornberger}, \citenamefont {Gerlich}, \citenamefont {Haslinger},
  \citenamefont {Nimmrichter},\ and\ \citenamefont
  {Arndt}}]{hornberger2012colloquium}%
  \BibitemOpen
  \bibfield  {author} {\bibinfo {author} {\bibfnamefont {Klaus}\ \bibnamefont
  {Hornberger}}, \bibinfo {author} {\bibfnamefont {Stefan}\ \bibnamefont
  {Gerlich}}, \bibinfo {author} {\bibfnamefont {Philipp}\ \bibnamefont
  {Haslinger}}, \bibinfo {author} {\bibfnamefont {Stefan}\ \bibnamefont
  {Nimmrichter}}, \ and\ \bibinfo {author} {\bibfnamefont {Markus}\
  \bibnamefont {Arndt}},\ }\bibfield  {title} {\enquote {\bibinfo {title}
  {Colloquium: Quantum interference of clusters and molecules},}\ }\href@noop
  {} {\bibfield  {journal} {\bibinfo  {journal} {Reviews of Modern Physics}\
  }\textbf {\bibinfo {volume} {84}},\ \bibinfo {pages} {157} (\bibinfo {year}
  {2012})}\BibitemShut {NoStop}%
\bibitem [{\citenamefont {Toro{\v{s}}}\ and\ \citenamefont
  {Bassi}(2018)}]{torovs2018bounds}%
  \BibitemOpen
  \bibfield  {author} {\bibinfo {author} {\bibfnamefont {Marko}\ \bibnamefont
  {Toro{\v{s}}}}\ and\ \bibinfo {author} {\bibfnamefont {Angelo}\ \bibnamefont
  {Bassi}},\ }\bibfield  {title} {\enquote {\bibinfo {title} {Bounds on quantum
  collapse models from matter-wave interferometry: calculational details},}\
  }\href@noop {} {\bibfield  {journal} {\bibinfo  {journal} {Journal of Physics
  A: Mathematical and Theoretical}\ }\textbf {\bibinfo {volume} {51}},\
  \bibinfo {pages} {115302} (\bibinfo {year} {2018})}\BibitemShut {NoStop}%
\bibitem [{\citenamefont {Toro{\v{s}}}\ \emph {et~al.}(2017)\citenamefont
  {Toro{\v{s}}}, \citenamefont {Gasbarri},\ and\ \citenamefont
  {Bassi}}]{torovs2017colored}%
  \BibitemOpen
  \bibfield  {author} {\bibinfo {author} {\bibfnamefont {Marko}\ \bibnamefont
  {Toro{\v{s}}}}, \bibinfo {author} {\bibfnamefont {Giulio}\ \bibnamefont
  {Gasbarri}}, \ and\ \bibinfo {author} {\bibfnamefont {Angelo}\ \bibnamefont
  {Bassi}},\ }\bibfield  {title} {\enquote {\bibinfo {title} {Colored and
  dissipative continuous spontaneous localization model and bounds from
  matter-wave interferometry},}\ }\href@noop {} {\bibfield  {journal} {\bibinfo
   {journal} {Physics Letters A}\ }\textbf {\bibinfo {volume} {381}},\ \bibinfo
  {pages} {3921--3927} (\bibinfo {year} {2017})}\BibitemShut {NoStop}%
\bibitem [{\citenamefont {Arndt}\ and\ \citenamefont
  {Hornberger}(2014)}]{arndt2014testing}%
  \BibitemOpen
  \bibfield  {author} {\bibinfo {author} {\bibfnamefont {Markus}\ \bibnamefont
  {Arndt}}\ and\ \bibinfo {author} {\bibfnamefont {Klaus}\ \bibnamefont
  {Hornberger}},\ }\bibfield  {title} {\enquote {\bibinfo {title} {Testing the
  limits of quantum mechanical superpositions},}\ }\href@noop {} {\bibfield
  {journal} {\bibinfo  {journal} {Nature Physics}\ }\textbf {\bibinfo {volume}
  {10}},\ \bibinfo {pages} {271} (\bibinfo {year} {2014})}\BibitemShut
  {NoStop}%
\bibitem [{\citenamefont {Bateman}\ \emph {et~al.}(2014)\citenamefont
  {Bateman}, \citenamefont {Nimmrichter}, \citenamefont {Hornberger},\ and\
  \citenamefont {Ulbricht}}]{bateman2014near}%
  \BibitemOpen
  \bibfield  {author} {\bibinfo {author} {\bibfnamefont {James}\ \bibnamefont
  {Bateman}}, \bibinfo {author} {\bibfnamefont {Stefan}\ \bibnamefont
  {Nimmrichter}}, \bibinfo {author} {\bibfnamefont {Klaus}\ \bibnamefont
  {Hornberger}}, \ and\ \bibinfo {author} {\bibfnamefont {Hendrik}\
  \bibnamefont {Ulbricht}},\ }\bibfield  {title} {\enquote {\bibinfo {title}
  {Near-field interferometry of a free-falling nanoparticle from a point-like
  source},}\ }\href@noop {} {\bibfield  {journal} {\bibinfo  {journal} {Nature
  communications}\ }\textbf {\bibinfo {volume} {5}},\ \bibinfo {pages} {4788}
  (\bibinfo {year} {2014})}\BibitemShut {NoStop}%
\bibitem [{\citenamefont {Kaltenbaek}\ \emph {et~al.}(2016)\citenamefont
  {Kaltenbaek}, \citenamefont {Aspelmeyer}, \citenamefont {Barker},
  \citenamefont {Bassi}, \citenamefont {Bateman}, \citenamefont {Bongs},
  \citenamefont {Bose}, \citenamefont {Braxmaier}, \citenamefont {Brukner},
  \citenamefont {Christophe} \emph {et~al.}}]{kaltenbaek2016macroscopic}%
  \BibitemOpen
  \bibfield  {author} {\bibinfo {author} {\bibfnamefont {Rainer}\ \bibnamefont
  {Kaltenbaek}}, \bibinfo {author} {\bibfnamefont {Markus}\ \bibnamefont
  {Aspelmeyer}}, \bibinfo {author} {\bibfnamefont {Peter~F}\ \bibnamefont
  {Barker}}, \bibinfo {author} {\bibfnamefont {Angelo}\ \bibnamefont {Bassi}},
  \bibinfo {author} {\bibfnamefont {James}\ \bibnamefont {Bateman}}, \bibinfo
  {author} {\bibfnamefont {Kai}\ \bibnamefont {Bongs}}, \bibinfo {author}
  {\bibfnamefont {Sougato}\ \bibnamefont {Bose}}, \bibinfo {author}
  {\bibfnamefont {Claus}\ \bibnamefont {Braxmaier}}, \bibinfo {author}
  {\bibfnamefont {{\v{C}}aslav}\ \bibnamefont {Brukner}}, \bibinfo {author}
  {\bibfnamefont {Bruno}\ \bibnamefont {Christophe}},  \emph {et~al.},\
  }\bibfield  {title} {\enquote {\bibinfo {title} {Macroscopic quantum
  resonators ({MAQRO}): 2015 update},}\ }\href@noop {} {\bibfield  {journal}
  {\bibinfo  {journal} {EPJ Quantum Technology}\ }\textbf {\bibinfo {volume}
  {3}},\ \bibinfo {pages} {5} (\bibinfo {year} {2016})}\BibitemShut {NoStop}%
\bibitem [{\citenamefont {Bassi}\ \emph {et~al.}(2017)\citenamefont {Bassi},
  \citenamefont {Gro{\ss}ardt},\ and\ \citenamefont
  {Ulbricht}}]{bassi2017gravitational}%
  \BibitemOpen
  \bibfield  {author} {\bibinfo {author} {\bibfnamefont {Angelo}\ \bibnamefont
  {Bassi}}, \bibinfo {author} {\bibfnamefont {Andr{\'e}}\ \bibnamefont
  {Gro{\ss}ardt}}, \ and\ \bibinfo {author} {\bibfnamefont {Hendrik}\
  \bibnamefont {Ulbricht}},\ }\bibfield  {title} {\enquote {\bibinfo {title}
  {Gravitational decoherence},}\ }\href@noop {} {\bibfield  {journal} {\bibinfo
   {journal} {Classical and Quantum Gravity}\ }\textbf {\bibinfo {volume}
  {34}},\ \bibinfo {pages} {193002} (\bibinfo {year} {2017})}\BibitemShut
  {NoStop}%
\bibitem [{\citenamefont {Storey}\ \emph {et~al.}(1992)\citenamefont {Storey},
  \citenamefont {Collett},\ and\ \citenamefont {Walls}}]{PhysRevLett.68.472}%
  \BibitemOpen
  \bibfield  {author} {\bibinfo {author} {\bibfnamefont {Pippa}\ \bibnamefont
  {Storey}}, \bibinfo {author} {\bibfnamefont {Matthew}\ \bibnamefont
  {Collett}}, \ and\ \bibinfo {author} {\bibfnamefont {Daniel}\ \bibnamefont
  {Walls}},\ }\bibfield  {title} {\enquote {\bibinfo {title}
  {Measurement-induced diffraction and interference of atoms},}\ }\href
  {\doibase 10.1103/PhysRevLett.68.472} {\bibfield  {journal} {\bibinfo
  {journal} {Phys. Rev. Lett.}\ }\textbf {\bibinfo {volume} {68}},\ \bibinfo
  {pages} {472--475} (\bibinfo {year} {1992})}\BibitemShut {NoStop}%
\bibitem [{\citenamefont {Abfalterer}\ \emph {et~al.}(1997)\citenamefont
  {Abfalterer}, \citenamefont {Keller}, \citenamefont {Bernet}, \citenamefont
  {Oberthaler}, \citenamefont {Schmiedmayer},\ and\ \citenamefont
  {Zeilinger}}]{PhysRevA.56.R4365}%
  \BibitemOpen
  \bibfield  {author} {\bibinfo {author} {\bibfnamefont {Roland}\ \bibnamefont
  {Abfalterer}}, \bibinfo {author} {\bibfnamefont {Claudia}\ \bibnamefont
  {Keller}}, \bibinfo {author} {\bibfnamefont {Stefan}\ \bibnamefont {Bernet}},
  \bibinfo {author} {\bibfnamefont {Markus~K.}\ \bibnamefont {Oberthaler}},
  \bibinfo {author} {\bibfnamefont {J\"org}\ \bibnamefont {Schmiedmayer}}, \
  and\ \bibinfo {author} {\bibfnamefont {Anton}\ \bibnamefont {Zeilinger}},\
  }\bibfield  {title} {\enquote {\bibinfo {title} {Nanometer definition of
  atomic beams with masks of light},}\ }\href {\doibase
  10.1103/PhysRevA.56.R4365} {\bibfield  {journal} {\bibinfo  {journal} {Phys.
  Rev. A}\ }\textbf {\bibinfo {volume} {56}},\ \bibinfo {pages} {R4365--R4368}
  (\bibinfo {year} {1997})}\BibitemShut {NoStop}%
\bibitem [{\citenamefont {Arndt}\ \emph {et~al.}(2014)\citenamefont {Arndt},
  \citenamefont {D{\"o}rre}, \citenamefont {Eibenberger}, \citenamefont
  {Haslinger}, \citenamefont {Rodewald}, \citenamefont {Hornberger},
  \citenamefont {Nimmrichter},\ and\ \citenamefont {Mayor}}]{arndt2014matter}%
  \BibitemOpen
  \bibfield  {author} {\bibinfo {author} {\bibfnamefont {M}~\bibnamefont
  {Arndt}}, \bibinfo {author} {\bibfnamefont {N}~\bibnamefont {D{\"o}rre}},
  \bibinfo {author} {\bibfnamefont {S}~\bibnamefont {Eibenberger}}, \bibinfo
  {author} {\bibfnamefont {P}~\bibnamefont {Haslinger}}, \bibinfo {author}
  {\bibfnamefont {J}~\bibnamefont {Rodewald}}, \bibinfo {author} {\bibfnamefont
  {K}~\bibnamefont {Hornberger}}, \bibinfo {author} {\bibfnamefont
  {S}~\bibnamefont {Nimmrichter}}, \ and\ \bibinfo {author} {\bibfnamefont
  {M}~\bibnamefont {Mayor}},\ }\href@noop {} {\emph {\bibinfo {title}
  {Matter-wave interferometry with composite quantum objects}}},\ Vol.\
  \bibinfo {volume} {188}\ (\bibinfo  {publisher} {IOS Press: Amsterdam, The
  Netherlands},\ \bibinfo {year} {2014})\BibitemShut {NoStop}%
\bibitem [{\citenamefont {Brezger}\ \emph {et~al.}(2003)\citenamefont
  {Brezger}, \citenamefont {Arndt},\ and\ \citenamefont
  {Zeilinger}}]{brezger2003concepts}%
  \BibitemOpen
  \bibfield  {author} {\bibinfo {author} {\bibfnamefont {Bj{\"o}rn}\
  \bibnamefont {Brezger}}, \bibinfo {author} {\bibfnamefont {Markus}\
  \bibnamefont {Arndt}}, \ and\ \bibinfo {author} {\bibfnamefont {Anton}\
  \bibnamefont {Zeilinger}},\ }\bibfield  {title} {\enquote {\bibinfo {title}
  {Concepts for near-field interferometers with large molecules},}\ }\href@noop
  {} {\bibfield  {journal} {\bibinfo  {journal} {Journal of Optics B: Quantum
  and Semiclassical Optics}\ }\textbf {\bibinfo {volume} {5}},\ \bibinfo
  {pages} {S82} (\bibinfo {year} {2003})}\BibitemShut {NoStop}%
\bibitem [{\citenamefont {Brezger}\ \emph {et~al.}(2002)\citenamefont
  {Brezger}, \citenamefont {Hackerm{\"u}ller}, \citenamefont {Uttenthaler},
  \citenamefont {Petschinka}, \citenamefont {Arndt},\ and\ \citenamefont
  {Zeilinger}}]{brezger2002matter}%
  \BibitemOpen
  \bibfield  {author} {\bibinfo {author} {\bibfnamefont {Bj{\"o}rn}\
  \bibnamefont {Brezger}}, \bibinfo {author} {\bibfnamefont {Lucia}\
  \bibnamefont {Hackerm{\"u}ller}}, \bibinfo {author} {\bibfnamefont {Stefan}\
  \bibnamefont {Uttenthaler}}, \bibinfo {author} {\bibfnamefont {Julia}\
  \bibnamefont {Petschinka}}, \bibinfo {author} {\bibfnamefont {Markus}\
  \bibnamefont {Arndt}}, \ and\ \bibinfo {author} {\bibfnamefont {Anton}\
  \bibnamefont {Zeilinger}},\ }\bibfield  {title} {\enquote {\bibinfo {title}
  {Matter-wave interferometer for large molecules},}\ }\href@noop {} {\bibfield
   {journal} {\bibinfo  {journal} {Physical review letters}\ }\textbf {\bibinfo
  {volume} {88}},\ \bibinfo {pages} {100404} (\bibinfo {year}
  {2002})}\BibitemShut {NoStop}%
\bibitem [{\citenamefont {Nairz}\ \emph {et~al.}(2003)\citenamefont {Nairz},
  \citenamefont {Arndt},\ and\ \citenamefont {Zeilinger}}]{nairz2003quantum}%
  \BibitemOpen
  \bibfield  {author} {\bibinfo {author} {\bibfnamefont {Olaf}\ \bibnamefont
  {Nairz}}, \bibinfo {author} {\bibfnamefont {Markus}\ \bibnamefont {Arndt}}, \
  and\ \bibinfo {author} {\bibfnamefont {Anton}\ \bibnamefont {Zeilinger}},\
  }\bibfield  {title} {\enquote {\bibinfo {title} {Quantum interference
  experiments with large molecules},}\ }\href@noop {} {\bibfield  {journal}
  {\bibinfo  {journal} {American Journal of Physics}\ }\textbf {\bibinfo
  {volume} {71}},\ \bibinfo {pages} {319--325} (\bibinfo {year}
  {2003})}\BibitemShut {NoStop}%
\bibitem [{\citenamefont {Gerlich}\ \emph {et~al.}(2011)\citenamefont
  {Gerlich}, \citenamefont {Eibenberger}, \citenamefont {Tomandl},
  \citenamefont {Nimmrichter}, \citenamefont {Hornberger}, \citenamefont
  {Fagan}, \citenamefont {T{\"u}xen}, \citenamefont {Mayor},\ and\
  \citenamefont {Arndt}}]{gerlich2011quantum}%
  \BibitemOpen
  \bibfield  {author} {\bibinfo {author} {\bibfnamefont {Stefan}\ \bibnamefont
  {Gerlich}}, \bibinfo {author} {\bibfnamefont {Sandra}\ \bibnamefont
  {Eibenberger}}, \bibinfo {author} {\bibfnamefont {Mathias}\ \bibnamefont
  {Tomandl}}, \bibinfo {author} {\bibfnamefont {Stefan}\ \bibnamefont
  {Nimmrichter}}, \bibinfo {author} {\bibfnamefont {Klaus}\ \bibnamefont
  {Hornberger}}, \bibinfo {author} {\bibfnamefont {Paul~J}\ \bibnamefont
  {Fagan}}, \bibinfo {author} {\bibfnamefont {Jens}\ \bibnamefont {T{\"u}xen}},
  \bibinfo {author} {\bibfnamefont {Marcel}\ \bibnamefont {Mayor}}, \ and\
  \bibinfo {author} {\bibfnamefont {Markus}\ \bibnamefont {Arndt}},\ }\bibfield
   {title} {\enquote {\bibinfo {title} {Quantum interference of large organic
  molecules},}\ }\href@noop {} {\bibfield  {journal} {\bibinfo  {journal}
  {Nature communications}\ }\textbf {\bibinfo {volume} {2}},\ \bibinfo {pages}
  {263} (\bibinfo {year} {2011})}\BibitemShut {NoStop}%
\bibitem [{\citenamefont {Pflanzer}\ \emph {et~al.}(2012)\citenamefont
  {Pflanzer}, \citenamefont {Romero-Isart},\ and\ \citenamefont
  {Cirac}}]{pflanzer2012master}%
  \BibitemOpen
  \bibfield  {author} {\bibinfo {author} {\bibfnamefont {Anika~C}\ \bibnamefont
  {Pflanzer}}, \bibinfo {author} {\bibfnamefont {Oriol}\ \bibnamefont
  {Romero-Isart}}, \ and\ \bibinfo {author} {\bibfnamefont {J~Ignacio}\
  \bibnamefont {Cirac}},\ }\bibfield  {title} {\enquote {\bibinfo {title}
  {Master-equation approach to optomechanics with arbitrary dielectrics},}\
  }\href@noop {} {\bibfield  {journal} {\bibinfo  {journal} {Physical Review
  A}\ }\textbf {\bibinfo {volume} {86}},\ \bibinfo {pages} {013802} (\bibinfo
  {year} {2012})}\BibitemShut {NoStop}%
\bibitem [{\citenamefont {Hornberger}\ \emph {et~al.}(2009)\citenamefont
  {Hornberger}, \citenamefont {Gerlich}, \citenamefont {Ulbricht},
  \citenamefont {Hackerm{\"u}ller}, \citenamefont {Nimmrichter}, \citenamefont
  {Goldt}, \citenamefont {Boltalina},\ and\ \citenamefont
  {Arndt}}]{hornberger2009theory}%
  \BibitemOpen
  \bibfield  {author} {\bibinfo {author} {\bibfnamefont {Klaus}\ \bibnamefont
  {Hornberger}}, \bibinfo {author} {\bibfnamefont {Stefan}\ \bibnamefont
  {Gerlich}}, \bibinfo {author} {\bibfnamefont {Hendrik}\ \bibnamefont
  {Ulbricht}}, \bibinfo {author} {\bibfnamefont {Lucia}\ \bibnamefont
  {Hackerm{\"u}ller}}, \bibinfo {author} {\bibfnamefont {Stefan}\ \bibnamefont
  {Nimmrichter}}, \bibinfo {author} {\bibfnamefont {Ilya~V}\ \bibnamefont
  {Goldt}}, \bibinfo {author} {\bibfnamefont {Olga}\ \bibnamefont {Boltalina}},
  \ and\ \bibinfo {author} {\bibfnamefont {Markus}\ \bibnamefont {Arndt}},\
  }\bibfield  {title} {\enquote {\bibinfo {title} {Theory and experimental
  verification of kapitza--dirac--talbot--lau interferometry},}\ }\href@noop {}
  {\bibfield  {journal} {\bibinfo  {journal} {New Journal of Physics}\ }\textbf
  {\bibinfo {volume} {11}},\ \bibinfo {pages} {043032} (\bibinfo {year}
  {2009})}\BibitemShut {NoStop}%
\bibitem [{\citenamefont {Hornberger}\ \emph {et~al.}(2004)\citenamefont
  {Hornberger}, \citenamefont {Sipe},\ and\ \citenamefont
  {Arndt}}]{PhysRevA.70.053608}%
  \BibitemOpen
  \bibfield  {author} {\bibinfo {author} {\bibfnamefont {Klaus}\ \bibnamefont
  {Hornberger}}, \bibinfo {author} {\bibfnamefont {John~E.}\ \bibnamefont
  {Sipe}}, \ and\ \bibinfo {author} {\bibfnamefont {Markus}\ \bibnamefont
  {Arndt}},\ }\bibfield  {title} {\enquote {\bibinfo {title} {Theory of
  decoherence in a matter wave talbot-lau interferometer},}\ }\href {\doibase
  10.1103/PhysRevA.70.053608} {\bibfield  {journal} {\bibinfo  {journal} {Phys.
  Rev. A}\ }\textbf {\bibinfo {volume} {70}},\ \bibinfo {pages} {053608}
  (\bibinfo {year} {2004})}\BibitemShut {NoStop}%
\bibitem [{\citenamefont {Nimmrichter}(2014)}]{nimmrichter2014macroscopic}%
  \BibitemOpen
  \bibfield  {author} {\bibinfo {author} {\bibfnamefont {Stefan}\ \bibnamefont
  {Nimmrichter}},\ }\href@noop {} {\emph {\bibinfo {title} {Macroscopic matter
  wave interferometry}}}\ (\bibinfo  {publisher} {Springer},\ \bibinfo {year}
  {2014})\BibitemShut {NoStop}%
\bibitem [{\citenamefont {Nimmrichter}\ and\ \citenamefont
  {Hornberger}(2008)}]{PhysRevA.78.023612}%
  \BibitemOpen
  \bibfield  {author} {\bibinfo {author} {\bibfnamefont {Stefan}\ \bibnamefont
  {Nimmrichter}}\ and\ \bibinfo {author} {\bibfnamefont {Klaus}\ \bibnamefont
  {Hornberger}},\ }\bibfield  {title} {\enquote {\bibinfo {title} {Theory of
  near-field matter-wave interference beyond the eikonal approximation},}\
  }\href {\doibase 10.1103/PhysRevA.78.023612} {\bibfield  {journal} {\bibinfo
  {journal} {Phys. Rev. A}\ }\textbf {\bibinfo {volume} {78}},\ \bibinfo
  {pages} {023612} (\bibinfo {year} {2008})}\BibitemShut {NoStop}%
\bibitem [{\citenamefont {Case}(2008)}]{case2008wigner}%
  \BibitemOpen
  \bibfield  {author} {\bibinfo {author} {\bibfnamefont {William~B}\
  \bibnamefont {Case}},\ }\bibfield  {title} {\enquote {\bibinfo {title}
  {Wigner functions and weyl transforms for pedestrians},}\ }\href@noop {}
  {\bibfield  {journal} {\bibinfo  {journal} {American Journal of Physics}\
  }\textbf {\bibinfo {volume} {76}},\ \bibinfo {pages} {937--946} (\bibinfo
  {year} {2008})}\BibitemShut {NoStop}%
\bibitem [{\citenamefont {Haslinger}\ \emph {et~al.}(2013)\citenamefont
  {Haslinger}, \citenamefont {D{\"o}rre}, \citenamefont {Geyer}, \citenamefont
  {Rodewald}, \citenamefont {Nimmrichter},\ and\ \citenamefont
  {Arndt}}]{haslinger2013universal}%
  \BibitemOpen
  \bibfield  {author} {\bibinfo {author} {\bibfnamefont {Philipp}\ \bibnamefont
  {Haslinger}}, \bibinfo {author} {\bibfnamefont {Nadine}\ \bibnamefont
  {D{\"o}rre}}, \bibinfo {author} {\bibfnamefont {Philipp}\ \bibnamefont
  {Geyer}}, \bibinfo {author} {\bibfnamefont {Jonas}\ \bibnamefont {Rodewald}},
  \bibinfo {author} {\bibfnamefont {Stefan}\ \bibnamefont {Nimmrichter}}, \
  and\ \bibinfo {author} {\bibfnamefont {Markus}\ \bibnamefont {Arndt}},\
  }\bibfield  {title} {\enquote {\bibinfo {title} {A universal matter-wave
  interferometer with optical ionization gratings in the time domain},}\
  }\href@noop {} {\bibfield  {journal} {\bibinfo  {journal} {Nature physics}\
  }\textbf {\bibinfo {volume} {9}},\ \bibinfo {pages} {144} (\bibinfo {year}
  {2013})}\BibitemShut {NoStop}%
\bibitem [{\citenamefont {Mie}(1908)}]{mie1908beitrage}%
  \BibitemOpen
  \bibfield  {author} {\bibinfo {author} {\bibfnamefont {Gustav}\ \bibnamefont
  {Mie}},\ }\bibfield  {title} {\enquote {\bibinfo {title} {Beitr{\"a}ge zur
  optik tr{\"u}ber medien, speziell kolloidaler metall{\"o}sungen},}\
  }\href@noop {} {\bibfield  {journal} {\bibinfo  {journal} {Annalen der
  physik}\ }\textbf {\bibinfo {volume} {330}},\ \bibinfo {pages} {377--445}
  (\bibinfo {year} {1908})}\BibitemShut {NoStop}%
\bibitem [{\citenamefont {Hulst}\ and\ \citenamefont {van~de
  Hulst}(1981)}]{hulst1981light}%
  \BibitemOpen
  \bibfield  {author} {\bibinfo {author} {\bibfnamefont {Hendrik~Christoffel}\
  \bibnamefont {Hulst}}\ and\ \bibinfo {author} {\bibfnamefont {Hendrik~C}\
  \bibnamefont {van~de Hulst}},\ }\href@noop {} {\emph {\bibinfo {title} {Light
  scattering by small particles}}}\ (\bibinfo  {publisher} {Courier
  Corporation},\ \bibinfo {year} {1981})\BibitemShut {NoStop}%
\bibitem [{\citenamefont {Bohren}\ and\ \citenamefont
  {Huffman}(2008)}]{bohren2008absorption}%
  \BibitemOpen
  \bibfield  {author} {\bibinfo {author} {\bibfnamefont {Craig~F}\ \bibnamefont
  {Bohren}}\ and\ \bibinfo {author} {\bibfnamefont {Donald~R}\ \bibnamefont
  {Huffman}},\ }\href@noop {} {\emph {\bibinfo {title} {Absorption and
  scattering of light by small particles}}}\ (\bibinfo  {publisher} {John Wiley
  \& Sons},\ \bibinfo {year} {2008})\BibitemShut {NoStop}%
\bibitem [{\citenamefont {Barton}\ \emph {et~al.}(1989)\citenamefont {Barton},
  \citenamefont {Alexander},\ and\ \citenamefont
  {Schaub}}]{barton1989theoretical}%
  \BibitemOpen
  \bibfield  {author} {\bibinfo {author} {\bibfnamefont {JP}~\bibnamefont
  {Barton}}, \bibinfo {author} {\bibfnamefont {DR}~\bibnamefont {Alexander}}, \
  and\ \bibinfo {author} {\bibfnamefont {SA}~\bibnamefont {Schaub}},\
  }\bibfield  {title} {\enquote {\bibinfo {title} {Theoretical determination of
  net radiation force and torque for a spherical particle illuminated by a
  focused laser beam},}\ }\href@noop {} {\bibfield  {journal} {\bibinfo
  {journal} {Journal of Applied Physics}\ }\textbf {\bibinfo {volume} {66}},\
  \bibinfo {pages} {4594--4602} (\bibinfo {year} {1989})}\BibitemShut {NoStop}%
\bibitem [{\citenamefont {Aspnes}\ and\ \citenamefont
  {Studna}(1983)}]{PhysRevB.27.985}%
  \BibitemOpen
  \bibfield  {author} {\bibinfo {author} {\bibfnamefont {D.~E.}\ \bibnamefont
  {Aspnes}}\ and\ \bibinfo {author} {\bibfnamefont {A.~A.}\ \bibnamefont
  {Studna}},\ }\bibfield  {title} {\enquote {\bibinfo {title} {Dielectric
  functions and optical parameters of si, ge, gap, gaas, gasb, inp, inas, and
  insb from 1.5 to 6.0 ev},}\ }\href {\doibase 10.1103/PhysRevB.27.985}
  {\bibfield  {journal} {\bibinfo  {journal} {Phys. Rev. B}\ }\textbf {\bibinfo
  {volume} {27}},\ \bibinfo {pages} {985--1009} (\bibinfo {year}
  {1983})}\BibitemShut {NoStop}%
\bibitem [{\citenamefont {Kitamura}\ \emph {et~al.}(2007)\citenamefont
  {Kitamura}, \citenamefont {Pilon},\ and\ \citenamefont
  {Jonasz}}]{kitamura2007optical}%
  \BibitemOpen
  \bibfield  {author} {\bibinfo {author} {\bibfnamefont {Rei}\ \bibnamefont
  {Kitamura}}, \bibinfo {author} {\bibfnamefont {Laurent}\ \bibnamefont
  {Pilon}}, \ and\ \bibinfo {author} {\bibfnamefont {Miroslaw}\ \bibnamefont
  {Jonasz}},\ }\bibfield  {title} {\enquote {\bibinfo {title} {Optical
  constants of silica glass from extreme ultraviolet to far infrared at near
  room temperature},}\ }\href@noop {} {\bibfield  {journal} {\bibinfo
  {journal} {Applied optics}\ }\textbf {\bibinfo {volume} {46}},\ \bibinfo
  {pages} {8118--8133} (\bibinfo {year} {2007})}\BibitemShut {NoStop}%
\bibitem [{\citenamefont {Pflanzer}(2013)}]{pflanzer2013optomechanics}%
  \BibitemOpen
  \bibfield  {author} {\bibinfo {author} {\bibfnamefont {Anika~C}\ \bibnamefont
  {Pflanzer}},\ }\emph {\bibinfo {title} {Optomechanics with Levitating
  Dielectrics: Theory and Protocols}},\ \href@noop {} {Ph.D. thesis},\ \bibinfo
   {school} {TU M{\"u}nchen M{\"u}nchen} (\bibinfo {year} {2013})\BibitemShut
  {NoStop}%
\bibitem [{\citenamefont {Erdelyi}\ \emph {et~al.}(1964)\citenamefont
  {Erdelyi}, \citenamefont {Magnus},\ and\ \citenamefont
  {Oberhettinger}}]{erdelyi1964m}%
  \BibitemOpen
  \bibfield  {author} {\bibinfo {author} {\bibfnamefont {A}~\bibnamefont
  {Erdelyi}}, \bibinfo {author} {\bibfnamefont {W}~\bibnamefont {Magnus}}, \
  and\ \bibinfo {author} {\bibfnamefont {F}~\bibnamefont {Oberhettinger}},\
  }\href@noop {} {\enquote {\bibinfo {title} {M. abramowitz and ia stegun,
  handbook of mathematical functions},}\ } (\bibinfo {year} {1964})\BibitemShut
  {NoStop}%
\bibitem [{\citenamefont {Jaynes}\ and\ \citenamefont
  {Cummings}(1963)}]{jaynes1963comparison}%
  \BibitemOpen
  \bibfield  {author} {\bibinfo {author} {\bibfnamefont {Edwin~T}\ \bibnamefont
  {Jaynes}}\ and\ \bibinfo {author} {\bibfnamefont {Frederick~W}\ \bibnamefont
  {Cummings}},\ }\bibfield  {title} {\enquote {\bibinfo {title} {Comparison of
  quantum and semiclassical radiation theories with application to the beam
  maser},}\ }\href@noop {} {\bibfield  {journal} {\bibinfo  {journal}
  {Proceedings of the IEEE}\ }\textbf {\bibinfo {volume} {51}},\ \bibinfo
  {pages} {89--109} (\bibinfo {year} {1963})}\BibitemShut {NoStop}%
\bibitem [{\citenamefont {Boukobza}\ and\ \citenamefont
  {Tannor}(2005)}]{boukobza2005entropy}%
  \BibitemOpen
  \bibfield  {author} {\bibinfo {author} {\bibfnamefont {E}~\bibnamefont
  {Boukobza}}\ and\ \bibinfo {author} {\bibfnamefont {DJ}~\bibnamefont
  {Tannor}},\ }\bibfield  {title} {\enquote {\bibinfo {title} {Entropy exchange
  and entanglement in the jaynes-cummings model},}\ }\href@noop {} {\bibfield
  {journal} {\bibinfo  {journal} {Physical Review A}\ }\textbf {\bibinfo
  {volume} {71}},\ \bibinfo {pages} {063821} (\bibinfo {year}
  {2005})}\BibitemShut {NoStop}%
\end{thebibliography}%

\appendix

\section{Mie Scattering}\label{AppendixA}
In this Appendix we collect the expressions used in the  text which derive from Mie scattering theory. We do not go into the details of the derivation, as we refer mostly to~\cite{bohren2008absorption,nimmrichter2014macroscopic} for an exhaustive treatment. Nonetheless, we explain how we compute the scattering amplitudes used in Sec.~\ref{IV}.

Mie theory serves to obtain an exact solution for the scattering of light off spherical homogeneous particles of arbitrary size. In a nutshell,  consider a plane electromagnetic wave $\mathbf{E}_0$ impinging on a homogeneous sphere. The latter will develop an internal field $\mathbf{E}_{\rm{int}}$ and modify the incident field adding a scattering component, in such a way that the external field is $\mathbf{E}_{\rm{ext}}=\mathbf{E}_{0}+\mathbf{E}_{s}$. From the symmetry of the problem, the internal, incident, and scattered fields can all be expanded in spherical harmonics. Then, by imposing boundary conditions on the transverse fields at the sphere surface (plus the fact that the internal field should be finite at $\mathbf{r}=0$), the scattered field can be related to the incident one, thus providing the scattering amplitudes and scattering and absorption cross-section(s).

The scattering coefficients, characterizing the scattering of a plane wave moving along the $z$-axis and linearly polarized in the $x$-direction, are given by
\begin{align}\label{scacoeff}
    & a_n=\frac{\sqrt{\epsilon}\psi_n(\sqrt{\epsilon}x)\psi_n'(x)-\psi_n(x)\psi_n'(\sqrt{\epsilon}x)}{\sqrt{\epsilon}\psi_n(\sqrt{\epsilon}x)\xi_n'(x)-\xi_n(x)\psi_n'(\sqrt{\epsilon}x)}\\
    & b_n=\frac{\psi_n(\sqrt{\epsilon}x)\psi_n'(x)-\sqrt{\epsilon}\psi_n(x)\psi_n'(\sqrt{\epsilon}x)}{\psi_n(\sqrt{\epsilon}x)\xi_n'(x)-\sqrt{\epsilon}\xi_n(x)\psi_n'(\sqrt{\epsilon}x)}.
\end{align}
where $x=kR$, and $\psi_{n},\xi_n$ are Riccati--Bessell functions, which are often expressed in terms of spherical Bessell $j$ functions and the spherical Henkel $h^{(1)}$ as 
\begin{align}
    & \psi_{n}(\rho)=\rho j_{n}(\rho)\\
    & \xi_n(\rho)=\rho h^{(1)}_n(\rho).
\end{align}
Here, as in the main text, $\epsilon$ is the relative permittivity of the sphere's material.

As should be clear from Sec.~\ref{IV}, in order to determine the scattering amplitudes to obtain the incoherent effect of scattering of grating photons, it is sufficient to consider the aforementioned case of an incident plane wave linearly polarized in the $x$-direction. The scattered field is related to the incident one via a vector scattering amplitudes $\mathbf{X}$
\[
\mathbf{E}_{s}\sim\frac{e^{i\mathbf{k}\cdot\mathbf{r}}}{kr}\mathbf{X}E_{0}.
\]
For spherical particles, the latter can be written in terms of the scalar scattering amplitude $f(\mathbf{k},\mathbf{k}')$ and the polarization direction of the scattered field $\hat{e}_s$ as
\begin{equation}
    \mathbf{X}=\sqrt{S_2^2\cos^2\phi+S_1^2\sin^2\phi}\,\hat{e}_s,
\end{equation}
where $\theta$ is the scattering angle and $\phi$ is the azimuthal angle with respect to the polarization direction. Thus the scattering amplitude reads
\begin{equation}
    f(\mathbf{k},\mathbf{k}')=\sqrt{S_2^2\cos^2\phi+S_1^2\sin^2\phi},
\end{equation}
where 
\begin{align}
    & S_1=\sum\frac{2n+1}{n(n+1)}(a_n \pi_n+b_n \tau_n)\\
    & S_2=\sum\frac{2n+1}{n(n+1)}(a_n \tau_n+b_n \pi_n).
\end{align}
The amplitude scattering matrix elements $S_{1,2}$ are given in terms of the scattering coefficients~\eqref{scacoeff} and the angular functions 
\begin{align}
    & \pi_n=\frac{P_n^1}{\sin\theta}=-\frac{dP_n(\cos\theta)}{d\theta}\frac{1}{\sin\theta}\\
    & \tau_n=\frac{dP^1_n}{d\theta}=\frac{d}{d\theta}\left(-\frac{dP_n(\cos\theta)}{d\theta}\right)
\end{align}
where $P_n(\cos\theta)$ are the Legendre polynomials of degree $n$.

Note that, in the Rayleigh limit ($x\ll 1$) the scattering coefficient $$a_1=-\frac{i 2x^3}{3}\frac{\epsilon-1}{\epsilon+2}+\mathcal{O}(x^5)$$ dominates, the scattering matrix elements become
\[ S^{\rm{(Ray)}}_1=\frac{3}{2}a_1,\,\,  S^{\rm{(Ray)}}_2=\frac{3}{2}a_1\cos\theta,\] and we recover the Rayleigh scattering result
\begin{equation}
     f(\mathbf{k},\mathbf{k}')= S^{\rm{(Ray)}}_1\sin\chi,
\end{equation}
where $\chi$ is the angle between the polarization direction of the incident light and the scattering direction.

Regarding Eq.~\eqref{shaub}, we need a slight extension of the Mie theory to account for interaction with standing waves. The way to obtain the final solution is the same as depicted before. We refer the reader to~\cite{nimmrichter2014macroscopic,barton1989theoretical} for the details of the calculation. Here we limit ourselves to reporting the expressions which appears in~\eqref{shaub}. The coefficients $A_{\ell m=\pm 1},\, B_{\ell m=\pm 1}$ are given by
\begin{align}
    & A_{\ell m}=\frac{i^{\ell+1} \sqrt{4 \pi  (2 \ell+1)}}{2\alpha ^2 \sqrt{l (l+1)}} m \zeta (\ell+1) \\
    & B_{\ell m}=\frac{i^\ell \sqrt{4 \pi  (2 \ell+1)}}{2\alpha ^2 \sqrt{l (l+1)}} \zeta (\ell) ,
\end{align}
where $\zeta(\ell)=\frac{1}{2} \left[(-1)^\ell \exp (-i k \text{z})+\exp (i k \text{z})\right]$ and $z$ represents the position of the center of mass of the sphere. The remaining coefficients, $a_{\ell m},\, b_{\ell m}$, appearing in Eq.~\eqref{shaub} are obtained by combining $A_{\ell m=\pm 1},\, B_{\ell m=\pm 1}$ with the scattering coefficients~\eqref{scacoeff}:
\begin{align}
    &a_{\ell m}=a_{\ell}A_{\ell m},\\
    &b_{\ell m}=b_{\ell}B_{\ell m}.
\end{align}

\end{document}